# TREELETS—AN ADAPTIVE MULTI-SCALE BASIS FOR SPARSE UNORDERED DATA

By Ann B. Lee,[2] Boaz Nadler[3] and Larry Wasserman[2]

*Carnegie Mellon University, Weizmann Institute of Science and Carnegie Mellon University*

In many modern applications, including analysis of gene expression and text documents, the data are noisy, high-dimensional, and unordered—with no particular meaning to the given order of the variables. Yet, successful learning is often possible due to sparsity: the fact that the data are typically redundant with underlying structures that can be represented by only a few features. In this paper we present *treelets*—a novel construction of multi-scale bases that extends wavelets to nonsmooth signals. The method is fully adaptive, as it returns a hierarchical tree and an orthonormal basis which *both* reflect the internal structure of the data. Treelets are especially well-suited as a dimensionality reduction and feature selection tool prior to regression and classification, in situations where sample sizes are small and the data are sparse with unknown groupings of correlated or collinear variables. The method is also simple to implement and analyze theoretically. Here we describe a variety of situations where treelets perform better than principal component analysis, as well as some common variable selection and cluster averaging schemes. We illustrate treelets on a blocked covariance model and on several data sets (hyperspectral image data, DNA microarray data, and internet advertisements) with highly complex dependencies between variables.

**1. Introduction.** For many modern data sets (e.g., DNA microarrays, financial and consumer data, text documents and internet web pages), the

Received May 2007; revised August 2007.
[1]Discussed in 10.1214/08-AOAS137A, 10.1214/08-AOAS137B, 10.1214/08-AOAS137C, 10.1214/08-AOAS137D, 10.1214/08-AOAS137E and 10.1214/08-AOAS137F; rejoinder at 10.1214/08-AOAS137REJ.
[2]Supported in part by NSF Grants CCF-0625879 and DMS-0707059.
[3]Supported by the Hana and Julius Rosen fund and by the William Z. and Eda Bess Novick Young Scientist fund.

collected data are *high-dimensional*, *noisy*, and *unordered*, with no particular meaning to the given order of the variables. In this paper we introduce a new methodology for the analysis of such data. We describe the theoretical properties of the method, and illustrate the proposed algorithm on hyperspectral image data, internet advertisements, and DNA microarray data. These data sets contain structure in the form of complex groupings of correlated variables. For example, the internet data include more than a thousand binary variables (various features of an image) and a couple of thousand observations (an image in an internet page). Some of the variables are exactly linearly related, while others are similar in more subtle ways. The DNA microarray data include the expression levels of several thousand genes but less than 100 samples (patients). Many sets of genes exhibit similar expression patterns across samples. The task in both cases is here classification. The results can therefore easily be compared with those of other classification algorithms. There is, however, a deeper underlying question that motivated our work: Is there a simple general methodology that, by construction, captures intrinsic localized structures, and that as a consequence improves inference and prediction of noisy, high-dimensional data when sample sizes are small? The method should be powerful enough to describe complex structures on multiple scales for unordered data, yet be simple enough to understand and analyze theoretically. Below we give some more background to this problem.

The key property that allows successful inference and prediction in high-dimensional settings is the notion of *sparsity*. Generally speaking, there are two main notions of sparsity. The first is sparsity of various quantities related either to the learning problem at hand or to the representation of the data in the original given variables. Examples include a sparse regression or classification vector [Tibshirani (1996)], and a sparse structure to the covariance or inverse covariance matrix of the given variables [Bickel and Levina (2008)]. The second notion is sparsity of the data *themselves*. Here we are referring to a situation where the data, despite their apparent high dimensionality, are highly redundant with underlying structures that can be represented by only a few features. Examples include data where many variables are approximately collinear or highly related, and data that lie on a nonlinear manifold [Belkin and Niyogi (2005), Coifman et al. (2005)].[1] While the two notions of sparsity are different, they are clearly related. In fact, a low intrinsic dimensionality of the data typically implies, for example, sparse regression or classification vectors, as well as low-rank covariance matrices. However, this relation may not be directly transparent, as the sparsity of these quantities sometimes becomes evident only in a different basis representation of the data.

---

[1] A referee pointed out that another issue with sparsity is that very high-dimensional spaces have very simple structure [Hall, Marron and Neeman (2005), Murtagh (2004), Ahn and Marron (2008)].

In either case, to take advantage of sparsity, one constrains the set of possible parameters of the problem. For the first kind of sparsity, two key tools are graphical models [Whittaker (2001)] that assume statistical dependence between specific variables, and *regularization methods* that penalize nonsparse solutions [Hastie, Tibshirani and Friedman (2001)]. Examples of such regularization methods are the lasso [Tibshirani (1996)], regularized covariance estimation methods [Bickel and Levina (2008), Levina and Zhu (2007)] and variable selection in high-dimensional graphs [Meinshausen and Bühlmann (2006)]. For the second type of sparsity, where the goal is to find a new set of coordinates or features of the data, two standard "variable transformation" methods are principal component analysis [Jolliffe (2002)] and wavelets [Ogden (1997)]. Each of these two methods has its own strengths and weaknesses which we briefly discuss here.

PCA has gained much popularity due to its simplicity and the unique property of providing a sequence of best linear approximations in a least squares sense. The method has two main limitations. First, PCA computes a global representation, where each basis vector is a linear combination of all the original variables. This makes it difficult to interpret the results and detect internal localized structures in the data. For example, in gene expression data, it may be difficult to detect small subsets of highly correlated genes. The second limitation is that PCA constructs an optimal linear representation of the noisy observations, but not necessarily of the (unknown) underlying noiseless data. When the number of variables $p$ is much larger than the number of observations $n$, the true underlying principal factors may be masked by the noise, yielding an inconsistent estimator in the joint limit $p, n \to \infty, p/n \to c$ [Johnstone and Lu (2008)]. Even for a finite sample size $n$, this property of PCA and other global methods (such as partial least squares and ridge regression) can lead to large prediction errors in regression and classification [Buckheit and Donoho (1995), Nadler and Coifman (2005b)]. Equation (25) in our paper, for example, gives an estimate of the finite-$n$ regression error for a linear mixture error-in-variables model.

In contrast to PCA, wavelet methods describe the data in terms of localized basis functions. The representations are multi-scale, and for smooth data, also sparse [Donoho and Johnstone (1995)]. Wavelets are used in many nonparametric statistics tasks, including regression and density estimation. In recent years wavelet expansions have also been combined with regularization methods to find regression vectors which are sparse in an a priori known wavelet basis [Candès and Tao (2007), Donoho and Elad (2003)]. The main limitation of wavelets is the implicit assumption of smoothness of the (noiseless) data as a function of its variables. In other words, standard wavelets are not suited for the analysis of unordered data. Thus, some work suggests first sorting the data, and then applying fixed wavelets to the reordered data [Murtagh, Starck and Berry (2000), Murtagh (2007)].

In this paper we propose an adaptive method for multi-scale representation and eigenanalysis of data where the variables can occur in any given order. We call the construction *treelets*, as the method is inspired by both hierarchical clustering trees and wavelets. The motivation for the treelets is two-fold: One goal is to find a "natural" system of coordinates that reflects *the underlying internal structure of the data* and that is robust to noise. A second goal is to improve the performance of conventional regression and classification techniques *in the "large p, small n"* regime by finding a reduced representation of the data prior to learning. We pay special attention to sparsity in the form of groupings of similar variables. Such low-dimensional structure naturally occurs in many data sets; for example, in DNA microarray data where genes sharing the same pathway can exhibit highly correlated expression patterns, and in the measured spectra of a chemical compound that is a linear mixture of certain simpler substances. Collinearity of variables is often a problem for a range of existing dimensionality reduction techniques—including least squares, and variable selection methods that do not take variable groupings into account.

The implementation of the treelet transform is similar to to the classical Jacobi method from numerical linear algebra [Golub and van Loan (1996)]. In our work we construct a data-driven multi-scale basis by applying a series of Jacobi rotations (PCA in two dimensions) to pairs of correlated variables. The final computed basis functions are orthogonal and supported on nested clusters in a hierarchical tree. As in standard PCA, we explore the covariance structure of the data but—unlike PCA—the analysis is *local* and *multi-scale*. As shown in Section 3.2.2 the treelet transform also has faster convergence properties than PCA. It is therefore more suitable as a feature extraction tool when sample sizes are small.

Other methods also relate to treelets. In recent years hierarchical clustering methods have been widely used for identifying diseases and groups of co-expressed genes [Eisen et al. (1998), Tibshirani et al. (1999)]. Many researchers are also developing algorithms that combine gene selection and gene grouping; see, for example, Hastie et al. (2001), Dettling and Bühlmann (2004), Zou and Hastie (2005) among others, and see Fraley and Raftery (2002) for a review of model-based clustering.

The novelty and contribution of our approach is the simultaneous construction of a data-driven multi-scale orthogonal basis *and* a hierarchical cluster tree. The introduction of a basis enables application of the well-developed machinery of orthonormal expansions, wavelets, and wavelet packets for nonparametric smoothing, data compression, and analysis of general unordered data. As with any orthonormal expansion, the expansion coefficients reflect the effective dimension of the data, as well as the significance of each coordinate. In our case, we even go one step further: *The basis functions themselves* contain information on the geometry of the data, while the corresponding expansion coefficients indicate their importance; see examples in Sections 4 and 5.

The treelet algorithm has some similarities to the *local* Karhunen–Loève Basis for smooth ordered data by Coifman and Saito (1996), where the basis functions are data-driven but the tree structure is fixed. Our paper is also related to recent independent work on the Haar wavelet transform of a dendrogram by Murtagh (2007). The latter paper also suggests basis functions on a data-driven cluster tree but uses fixed wavelets on a pre-computed dendrogram. The treelet algorithm offers the advantages of *both* approaches, as it incorporates adaptive basis functions as well as a data-driven tree structure. As shown in this paper, this unifying property turns out to be of key importance for statistical inference and prediction: The adaptive tree structure allows analysis of unordered data. The adaptive treelet functions lead to results that reflect the internal localized structure of the data, and that are stable to noise. In particular, when the data contain subsets of co-varying variables, the computed basis is sparse, with the dominant basis functions effectively serving as indicator functions of the hidden groups. For more complex structure, as illustrated on real data sets, our method returns "softer," continuous-valued loading functions. In classification problems, the treelet functions with the most discriminant power often compute differences between groups of variables.

The organization of the paper is as follows: In Section 2 we describe the treelet algorithm. In Section 3 we examine its theoretical properties. The analysis includes the general large-sample properties of treelets, as well as a specific covariance model with block structure. In Section 4 we discuss the performance of the treelet method on a linear mixture error-in-variable model and give a few illustrative examples of its use in data representation and regression. Finally, in Section 5 we apply our method to classification of hyperspectral data, internet advertisements, and gene expression arrays.

A preliminary version of this paper was presented at AISTATS-07 [Lee and Nadler (2007)].

**2. The treelet transform.** In many modern data sets the data are not only high-dimensional but also redundant with many variables related to each other. Hierarchical clustering algorithms [Jain, Murty and Flynn (1999), Xu and Wunsch (2005)] are often used for the organization and grouping of the variables of such data sets. These methods offer an easily interpretable description of the data structure in terms of a dendrogram, and only require the user to specify a measure of similarity between groups of observations or variables. So called agglomerative hierarchical methods start at the bottom of the tree and, at each level, merge the two groups with highest inter-group similarity into one larger cluster. The novelty of the proposed treelet algorithm is in constructing not only clusters or groupings of variables, but also functions on the data. More specifically, we construct a *multi-scale orthonormal basis on a hierarchical tree.* As in standard multi-resolution analysis [Mallat (1998)], the treelet algorithm provides a set of "scaling functions" defined on nested subspaces $V_0 \supset V_1 \supset \cdots \supset V_L$, and a set of orthogonal "detail functions" defined on residual spaces $\{W_\ell\}_{\ell=1}^{L}$, where $V_\ell \oplus W_\ell = V_{\ell-1}$. The treelet decomposition scheme represents a *multi-resolution transform*, but technically speaking, not a wavelet transform. (In terms of the tiling of "time-frequency" space, the method is more similar to local cosine transforms, which divide the time axis in intervals of varying sizes.) The details of the treelet algorithm are in Section 2.1.

In this paper we measure the similarity $M_{ij}$ between two variables $s_i$ and $s_j$ with the correlation coefficient

$$\rho_{ij} = \frac{\Sigma_{ij}}{\sqrt{\Sigma_{ii}\Sigma_{jj}}}, \tag{1}$$

where $\Sigma_{ij} = \mathbb{E}[(s_i - \mathbb{E}s_i)(s_j - \mathbb{E}s_j)]$ is the usual covariance. Other information-theoretic or graph-theoretic similarity measures are also possible. For some applications, one may want to use absolute values of correlation coefficients, or a weighted sum of covariances and correlations as in $M_{ij} = |\rho_{ij}| + \lambda|\Sigma_{ij}|$, where the parameter $\lambda$ is a nonnegative number.

2.1. *The algorithm: Jacobi rotations on pairs of similar variables.* The treelet algorithm is inspired by the classical Jacobi method for computing eigenvalues of a matrix [Golub and van Loan (1996)]. There are also some similarities with the Grand Tour [Asimov (1985)], a visualization tool for viewing multidimensional data through a sequence of orthogonal projections. The main difference from Jacobi's method—and the reason why the treelet transform, in general, returns an orthonormal basis *different* from standard PCA—is that treelets are constructed on a hierarchical tree.

The idea is simple. At each level of the tree, we group together the most similar variables and replace them by a coarse-grained "sum variable" and a residual "difference variable." The new variables are computed by a local PCA (or Jacobi rotation) in two dimensions. Unlike Jacobi's original method, difference variables are stored, and only sum variables are processed at higher levels of the tree. Hence, the multi-resolution analysis (MRA) interpretation. The details of the algorithm are as follows:

- At level $\ell = 0$ (the bottom of the tree), each observation or "signal" x is represented by the original variables $\mathbf{x}^{(0)} = [s_{0,1}, \ldots, s_{0,p}]^T$, where $s_{0,k} = x_k$. Associate to these coordinates, the Dirac basis, $B_0 = [\phi_{0,1}, \phi_{0,2}, \ldots, \phi_{0,p}]$, where $B_0$ is the $p \times p$ identity matrix. Compute the sample covariance and similarity matrices $\hat{\Sigma}^{(0)}$ and $\hat{M}^{(0)}$. Initialize the set of "sum variables," $\mathcal{S} = \{1, 2, \ldots, p\}$.
- Repeat for $\ell = 1, \ldots, L$:

  1. **Find the two most similar sum variables according to the similarity matrix $\hat{M}^{(\ell-1)}$.** Let

  $$(\alpha, \beta) = \arg\max_{i,j \in \mathcal{S}} \hat{M}_{ij}^{(\ell-1)}, \qquad (2)$$

  where $i < j$, and maximization is only over pairs of sum variables that belong to the set $\mathcal{S}$. As in standard wavelet analysis, difference variables (defined in step 3) are not processed.

  2. **Perform a local PCA on this pair.** Find a Jacobi rotation matrix

  $$J(\alpha, \beta, \theta_\ell) = \begin{bmatrix} 1 & \cdots & 0 & & 0 & \cdots & 0 \\ \vdots & \ddots & \vdots & & \vdots & & \vdots \\ 0 & \cdots & c & \cdots & -s & \cdots & 0 \\ \vdots & & \vdots & \ddots & \vdots & & \vdots \\ 0 & \cdots & s & \cdots & c & \cdots & 0 \\ \vdots & & \vdots & & \vdots & \ddots & \vdots \\ 0 & \cdots & 0 & \cdots & 0 & \cdots & 1 \end{bmatrix}, \qquad (3)$$

  where $c = \cos(\theta_\ell)$ and $s = \sin(\theta_\ell)$, that decorrelates $x_\alpha$ and $x_\beta$; more specifically, find a rotation angle $\theta_\ell$ such that $|\theta_\ell| \leq \pi/4$ and $\hat{\Sigma}_{\alpha\beta}^{(\ell)} = \hat{\Sigma}_{\beta\alpha}^{(\ell)} = 0$, where $\hat{\Sigma}^{(\ell)} = J^T \hat{\Sigma}^{(\ell-1)} J$. This transformation corresponds to a change of basis $B_\ell = B_{\ell-1} J$, and new coordinates $\mathbf{x}^{(\ell)} = J^T \mathbf{x}^{(\ell-1)}$. Update the similarity matrix $\hat{M}^{(\ell)}$ accordingly.

  3. **Multi-resolution analysis.** For ease of notation, assume that $\hat{\Sigma}_{\alpha\alpha}^{(\ell)} \geq \hat{\Sigma}_{\beta\beta}^{(\ell)}$ after the Jacobi rotation, where the indices $\alpha$ and $\beta$ correspond to the first and second principal components, respectively. Define the sum and difference variables at level $\ell$ as $s_\ell = x_\alpha^{(\ell)}$ and $d_\ell = x_\beta^{(\ell)}$. Similarly, define the scaling and detail functions $\phi_\ell$ and $\psi_\ell$ as columns $\alpha$ and $\beta$ of the basis matrix $B_\ell$. Remove the difference variable from the set of sum variables, $\mathcal{S} = \mathcal{S} \setminus \{\beta\}$. At level $\ell$, we have the orthonormal *treelet decomposition*

  $$\mathbf{x} = \sum_{i=1}^{p-\ell} s_{\ell,i} \phi_{\ell,i} + \sum_{i=1}^{\ell} d_i \psi_i, \qquad (4)$$

  where the new set of scaling vectors $\{\phi_{\ell,i}\}_{i=1}^{p-\ell}$ is the union of the vector $\phi_\ell$ and the scaling vectors $\{\phi_{\ell-1,j}\}_{j \neq \alpha, \beta}$ from the previous level, and the new coarse-grained sum variables $\{s_{\ell,i}\}_{i=1}^{p-\ell}$ are the projections of the original data onto these vectors. As in standard multi-resolution analysis, the first sum is the coarse-grained representation of the signal, while the second sum captures the residuals at different scales.

The output of the algorithm can be summarized in terms of a hierarchical tree with a height $L \leq p-1$ and an ordered set of rotations and pairs of indices, $\{(\theta_\ell, \alpha_\ell, \beta_\ell)\}_{\ell=1}^{L}$. Figure 1 (left) shows an example of a treelet construction for a "signal" of length $p = 5$, with the data representations $\mathbf{x}^{(\ell)}$ at the different levels of the tree shown on the right. The $s$-components (projections in the main principal directions) represent coarse-grained "sums." We associate these variables to the nodes in the cluster tree. Similarly, the $d$-components (projections in the orthogonal directions) represent "differences" between node representations at two consecutive levels in the tree. For example, in the figure $d_1 \psi_1 = (s_{0,1}\phi_{0,1} + s_{0,2}\phi_{0,2}) - s_1 \phi_{1,1}$.

We now briefly consider the complexity of the treelet algorithm on a general data set with $n$ observations and $p$ variables. For a naive implementation with an exhaustive search for the optimal pair $(\alpha, \beta)$ in Equation 2, the overall complexity is $m + O(Lp^2)$ operations, where $m = O(\min(np^2, pn^2))$ is the cost of computing the sample covariance matrix by singular value decomposition, and $L$ is the height of the tree. However, by storing the similarity matrices $\hat{\Sigma}^{(0)}$ and $\hat{M}^{(0)}$ and keeping track of their local changes, the complexity can be further reduced to $m + O(Lp)$. In other words, the computational cost is comparable to hierarchical clustering algorithms.

2.2. *Selecting the height $L$ of the tree and a "best K-basis."* The default choice of the treelet transform is a maximum height tree with $L = p - 1$;

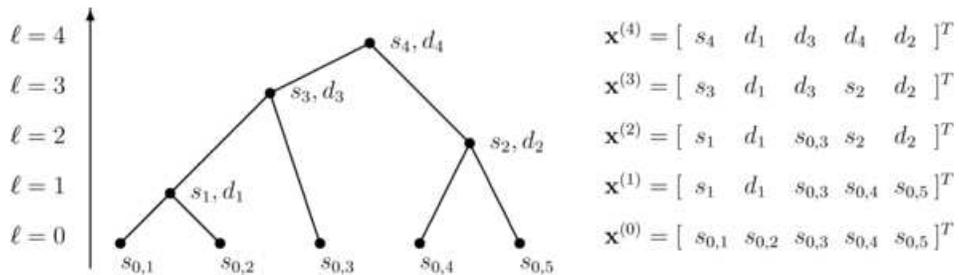

FIG. 1. (Left) *A toy example of a hierarchical tree for data of dimension $p = 5$. At $\ell = 0$, the signal is represented by the original $p$ variables. At each successive level $\ell = 1, 2, \ldots, p-1$, the two most similar sum variables are combined and replaced by the sum and difference variables $s_\ell, d_\ell$ corresponding to the first and second local principal components.* (Right) *Signal representation $\mathbf{x}^{(\ell)}$ at different levels. The $s$- and $d$-coordinates represent projections along scaling and detail functions in a multi-scale treelet decomposition. Each such representation is associated with an orthogonal basis in $\mathbb{R}^p$ that captures the local eigenstructure of the data.*

see examples in Sections 5.1 and 5.3. This choice leads to a fully *parameter-free* decomposition of the data and is also faithful to the idea of a multi-resolution analysis. For more complexity, one can alternatively also choose any of the orthonormal (ON) bases at levels $\ell < p - 1$ of the tree. The data are then represented by coarse-grained sum variables for a set of clusters in the tree, and difference variables that describe the finer details. In principle, any of the standard techniques in hierarchical clustering can be used in deciding when to stop "merging" clusters (e.g., use a preset threshold value for the similarity measure, or use hypothesis testing for homogeneity of clusters, etc.). In this work we propose a rather different method that is inspired by the *best basis paradigm* [Coifman and Wickerhauser (1992), Saito and Coifman (1995)] in wavelet signal processing. This approach directly addresses the question of how well one can capture information in the data.

Consider IID data $\mathbf{x}^1, \ldots, \mathbf{x}^n$, where $\mathbf{x}^i \in \mathbb{R}^p$ is a $p$-dimensional random vector. Denote the candidate ON bases by $B_0, \ldots, B_{p-1}$, where $B_\ell$ is the basis at level $\ell$ in the tree. Suppose now that we are interested in finding the "best" $K$-dimensional treelet representation for data representation and compression, where the dimension $K < p$ has been determined in advance. It then makes sense to use a scoring criterion that measures the percentage of explained variance for the chosen coordinates. Thus, we propose the following greedy scoring and selection approach:

For a given orthonormal basis $B = (\mathbf{w}_1, \ldots, \mathbf{w}_p)$, assign a *normalized energy score* $\mathcal{E}$ to each vector $\mathbf{w}_i$ according to

$$\mathcal{E}(\mathbf{w}_i) = \frac{\mathbb{E}\{|\mathbf{w}_i \cdot \mathbf{x}|^2\}}{\mathbb{E}\{\|\mathbf{x}\|^2\}}. \tag{5}$$

The corresponding sample estimate is $\hat{\mathcal{E}}(\mathbf{w}) = \frac{\sum_{j=1}^n |\mathbf{w}_i \cdot \mathbf{x}^j|^2}{\sum_{j=1}^n \|\mathbf{x}^j\|^2}$. Sort the vectors according to decreasing energy, $\mathbf{w}_{(1)}, \ldots, \mathbf{w}_{(p)}$, and define the score $\Gamma_K$ of the basis $B$ by summing the $K$ largest terms, that is, let $\Gamma_K(B) \equiv \sum_{i=1}^K \mathcal{E}(\mathbf{w}_i)$. The *best K-basis* is the treelet basis with the highest score

$$B_L = \arg\max_{B_\ell:\, 0 \leq \ell \leq p-1} \Gamma_K(B_\ell). \tag{6}$$

It is the basis that best compresses the data with only $K$ components. In case of degeneracies, we choose the coordinate system with the smallest $\ell$. Furthermore, to estimate the score $\Gamma_K$ for a particular data set, we use cross-validation (CV); that is, the treelets are constructed using subsets of the original data set and the score is computed on independent test sets to avoid overfitting. Both theoretical calculations (Section 3.2) and simulations (Section 4.1) indicate that an energy-based measure is useful for detecting natural groupings of variables in data. Many alternative measures (e.g., Fisher's discriminant score, classification error rates, entropy, and other sparsity measures) can also be used. For the classification problem in Section 5.1, for example, we define a discriminant score that measures how well a coordinate separates data from different classes.

## 3. Theory.

3.1. *Large sample properties of the treelet transform.* In this section we examine the large sample properties of treelets. We introduce a more general definition of consistency that takes into account the fact that the treelet operator (based on correlation coefficients) is multi-valued, and study the method under the stated conditions. We also describe a bootstrap algorithm for quantifying the stability of the algorithm in practical applications. The details are as follows.

First some notation and definitions: Let $T(\Sigma) = J^T \Sigma J$ denote the covariance matrix after one step of the treelet algorithm when starting with covariance matrix $\Sigma$. Let $T^\ell(\Sigma)$ denote the covariance matrix after $\ell$ steps of the treelet algorithm. Thus, $T^\ell = T \circ \cdots \circ T$ corresponds to $T$ applied $\ell$ times. Define $\|A\|_\infty = \max_{j,k} |A_{jk}|$ and let

$$(7) \qquad \mathcal{T}_n(\Sigma, \delta_n) = \bigcup_{\|\Lambda - \Sigma\|_\infty \leq \delta_n} T(\Lambda).$$

Define $\mathcal{T}_n^1(\Sigma, \delta_n) = \mathcal{T}_n(\Sigma, \delta_n)$, and

$$(8) \qquad \mathcal{T}_n^\ell(\Sigma, \delta_n) = \bigcup_{\Lambda \in \mathcal{T}_n^{\ell-1}} T(\Lambda), \qquad \ell \geq 2.$$

Let $\hat{\Sigma}_n$ denote the sample covariance matrix. We make the following assumptions:
(A1) Assume that $\mathbf{x}$ has finite variance and satisfies one of the following three assumptions: (a) each $x_j$ is bounded or (b) $\mathbf{x}$ is multivariate normal or (c) there exist $M$ and $s$ such that $\mathbb{E}(|x_j x_k|^q) \leq q! M^{q-2} s/2$ for all $q \geq 2$.
(A2) The dimension $p_n$ satisfies $p_n \leq n^c$ for some $c > 0$.

THEOREM 1. *Suppose that* (A1) *and* (A2) *hold. Let* $\delta_n = K\sqrt{\log n/n}$, *where* $K > 2c$. *Then, as* $n, p_n \to \infty$,

$$(9) \qquad \mathbb{P}(T^\ell(\hat{\Sigma}_n) \in \mathcal{T}_n^\ell(\Sigma, \delta_n), \ell = 1, \ldots, p_n) \to 1.$$

Some discussion is in order. The result says that $T^\ell(\hat{\Sigma}_n)$ is not too far from $T^\ell(\Lambda)$ for some $\Lambda$ close to $\Sigma$. It would perhaps be more satisfying to have a result that says that $\|T^\ell(\Sigma) - T^\ell(\hat{\Sigma})\|_\infty$ converges to 0. This would be possible if one used covariances to measure similarity, but not in the case of correlation coefficients.

For example, it is easy to construct a covariance matrix $\Sigma$ with the following properties:

1. $\rho_{12}$ is the largest off-diagonal correlation,
2. $\rho_{34}$ is nearly equal to $\rho_{12}$,
3. the $2 \times 2$ submatrix of $\Sigma$ corresponding to $x_1$ and $x_2$ is very different than the $2 \times 2$ submatrix of $\Sigma$ corresponding to $x_3$ and $x_4$.

In this case there is a nontrivial probability that $\hat{\rho}_{34} > \hat{\rho}_{12}$ due to sample fluctuations. Therefore $T(\Sigma)$ performs a rotation on the first two coordinates, while $T(\hat{\Sigma})$ performs a rotation on the third and fourth coordinates. Since the two corresponding submatrices are quite different, the two rotations will be quite different. Hence, $T(\Sigma)$ can be quite different from $T(\hat{\Sigma})$. This does not pose any problem since inferring $T(\Sigma)$ is not the goal. Under the stated conditions, we would consider both $T(\Sigma)$ and $T(\hat{\Sigma})$ to be reasonable transformations. We examine the details and include the proof of Theorem 1 in Appendix A.1.

Because $T(\Sigma_1)$ and $T(\Sigma_2)$ can be quite different even when the matrices $\Sigma_1$ and $\Sigma_2$ are close, it might be of interest to study the stability of $T(\hat{\Sigma}_n)$. This can be done using the *bootstrap*. Construct $B$ bootstrap replications of the data and corresponding sample covariance matrices $\hat{\Sigma}^*_{n,1}, \ldots, \hat{\Sigma}^*_{n,B}$. Let $\delta_n = J_n^{-1}(1-\alpha)$, where $J_n$ is the empirical distribution function of $\{\|\hat{\Sigma}^*_{n,b} - \hat{\Sigma}_n\|_\infty, b=1,\ldots,B\}$ and $\alpha$ is the confidence level. If $F$ has finite fourth moments and $p$ is fixed, then it follows from Corollary 1 of Beran and Srivastava (1985) that

$$\lim_{n \to \infty} \mathbb{P}_F(\Sigma \in C_n) = 1 - \alpha,$$

where $C_n = \{\Lambda : \|\Lambda - \hat{\Sigma}_n\|_\infty \leq \delta_n\}$. Let

$$A_n = \{T(\Lambda) : \Lambda \in C_n\}.$$

It follows that $\mathbb{P}(T(\Sigma) \in A_n) \to 1 - \alpha$. The set $A_n$ can be approximated by applying $T$ to all $\hat{\Sigma}^*_{n,b}$ for which $\|\hat{\Sigma}^*_{n,b} - \hat{\Sigma}_n\|_\infty < \delta_n$. In Section 4.1 (Figure 3) we use the bootstrap method to estimate confidence sets for treelets.

### 3.2. Treelets on covariance matrices with block structures.

#### 3.2.1. An exact analysis in the limit $n \to \infty$.
Many real life data sets, including gene arrays, consumer data sets, and word-documents, display covariance matrices with approximate block structures. The treelet transform is especially well suited for representing and analyzing such data—even for noisy data and small sample sizes.

Here we show that treelets provide a sparse representation when covariance matrices have inherent block structures, and that the loading functions themselves contain information about the inherent groupings. We consider an ideal situation where variables within the same group are collinear, and variables from different groups are weakly correlated. All calculations are exact and computed in the limit of the sample size $n \to \infty$. An analysis of convergence rates later appears in Section 3.2.2.

We begin by analyzing treelets on $p$ random variables that are indistinguishable with respect to their second-order statistics. We show that the treelet algorithm returns scaling functions that are *constant on groups of indistinguishable variables*. In particular, the scaling function on the full set of variables in a block is a constant function. Effectively, this function serves as an indicator function of a (sometimes hidden) set of similar variables in data. These results, as well as the follow-up main results in Theorem 2 and Corollary 1, are due to the *fully adaptive* nature of the treelet algorithm—a property that sets treelets apart from methods that use fixed wavelets on a dendrogram [Murtagh (2007)], or adaptive basis functions on fixed trees [Coifman and Saito (1996)]; see Remark 2 for a concrete example.

LEMMA 1. *Assume that* $\mathbf{x} = (x_1, x_2, \ldots, x_p)^T$ *is a random vector with distribution $F$, mean 0, and covariance matrix $\Sigma = \sigma_1^2 1_{p \times p}$, where $1_{p \times p}$ denotes a $p \times p$ matrix with all entries equal to 1. Then, at any level $1 \le \ell \le p-1$ of the tree, the treelet operator $T^\ell$ (defined in Section 3.1) returns (for the population covariance matrix $\Sigma$) an orthogonal decomposition*

$$(10) \qquad T^\ell(\Sigma) = \sum_{i=1}^{p-\ell} s_{\ell,i} \phi_{\ell,i} + \sum_{i=1}^{\ell} d_i \psi_i,$$

*with sum variables* $s_{\ell,i} = \frac{1}{\sqrt{|\mathcal{A}_{\ell,i}|}} \sum_{j \in \mathcal{A}_{\ell,i}} x_j$ *and scaling functions* $\phi_{\ell,i} = \frac{1}{\sqrt{|\mathcal{A}_{\ell,i}|}} \times I_{s_{\ell,i}}$, *which are defined on disjoint index subsets* $\mathcal{A}_{\ell,i} \subseteq \{1, \ldots, p\}$ $(i = 1, \ldots, p-\ell)$ *with lengths* $|\mathcal{A}_{\ell,i}|$ *and* $\sum_{i=1}^{p-\ell} |\mathcal{A}_{\ell,i}| = p$. *The expansion coefficients have variances* $\mathbb{V}\{s_{\ell,i}\} = |\mathcal{A}_{\ell,i}| \sigma_1^2$, *and* $\mathbb{V}\{d_i\} = 0$. *In particular, for* $\ell = p-1$,

$$(11) \qquad T^{p-1}(\Sigma) = s\phi + \sum_{i=1}^{p-1} d_i \psi_i,$$

*where* $s = \frac{1}{\sqrt{p}}(x_1 + \cdots + x_p)$ *and* $\phi = \frac{1}{\sqrt{p}}[1 \ldots 1]^T$.

REMARK 1. Uncorrelated additive noise in $(x_1, x_2, \ldots, x_p)$ adds a diagonal perturbation to the $2 \times 2$ covariance matrices $\Sigma^{(\ell)}$, which are computed at each level in the tree [see (35)]. Such noise may affect the order in which variables are grouped, but the asymptotic results of the lemma remain the same.

REMARK 2. The treelet algorithm is robust to noise because it computes *data-driven* rotations on variables. On the other hand, methods that use fixed transformations on pre-computed trees are often highly sensitive to noise, yielding inconsistent results. Consider, for example, a set of four statistically indistinguishable variables $\{x_1, x_2, x_3, x_4\}$, and compare treelets to a Haar wavelet transform on a data-driven dendrogram [Murtagh (2004)]. The two methods return the same results if the variables are merged in the order $\{\{x_1, x_2\}, \{x_3, x_4\}\}$; that is, $s = \frac{1}{2}(x_1 + x_2 + x_3 + x_4)$ and $\phi = \frac{1}{2}[1, 1, 1, 1]^T$. Now, a different realization of the noise may lead to the order $\{\{\{x_1, x_2\}, x_4\}, x_3\}$. A *fixed* rotation angle of $\pi/4$ (as in Haar wavelets) would then return the sum variable $s^{\text{Haar}} = \frac{1}{\sqrt{2}}(\frac{1}{\sqrt{2}}(\frac{1}{\sqrt{2}}(x_1 + x_2) + x_4) + x_3)$ and scaling function $\phi^{\text{Haar}} = [\frac{1}{2\sqrt{2}}, \frac{1}{2\sqrt{2}}, \frac{1}{\sqrt{2}}, \frac{1}{2}]^T$.

Next we consider data where the covariance matrix is a $K \times K$ block matrix with white noise added to the original variables. The following main result states that, *if variables from different blocks are weakly correlated and the noise level is relatively small, then the $K$ maximum variance scaling functions are constant on each block* (see Figure 2 in Section 4 for an example). We make this precise by giving a sufficient condition [equation (13)] in terms of the noise level, and within-block and between-block correlations of the original data. For illustrative purposes, we have reordered the variables. A $p \times p$ identity matrix is denoted by $I_p$, and a $p_i \times p_j$ matrix with all entries equal to 1 is denoted by $1_{p_i \times p_j}$.

THEOREM 2. *Assume that* $\mathbf{x} = (x_1, x_2, \ldots, x_p)^T$ *is a random vector with distribution $F$, mean 0 and covariance matrix* $\Sigma = C + \sigma^2 I_p$, *where $\sigma^2$ represents the variance of white noise in each variable and*

$$
(12) \qquad C = \begin{pmatrix} C_{11} & C_{12} & \ldots & C_{1K} \\ C_{12} & C_{22} & \ldots & C_{2K} \\ \vdots & \vdots & \ddots & \vdots \\ C_{1K} & C_{2K} & \ldots & C_{KK} \end{pmatrix}
$$

*is a $K \times K$ block matrix with "within-block" covariance matrices* $C_{kk} = \sigma_k^2 1_{p_k \times p_k}$ ($k = 1, \ldots, K$) *and "between-block" covariance matrices* $C_{ij} = \sigma_{ij} 1_{p_i \times p_j}$ ($i, j = 1, \ldots, K; i \neq j$). *If*

$$
(13) \qquad \max_{1 \leq i,j \leq K} \left( \frac{\sigma_{ij}}{\sigma_i \sigma_j} \right) < \frac{1}{\sqrt{1 + 3\max(\delta^2, \delta^4)}},
$$

*where* $\delta = \frac{\sigma}{\min_k \sigma_k}$, *then the treelet decomposition at level* $\ell = p - K$ *has the form*

$$
(14) \qquad T^{p-K}(\Sigma) = \sum_{k=1}^{K} s_k \phi_k + \sum_{i=1}^{p-K} d_i \psi_i,
$$

where $s_k = \frac{1}{\sqrt{p_k}} \sum_{j \in \mathcal{B}_k} x_j$, $\phi_k = \frac{1}{\sqrt{p_k}} I_{\mathcal{B}_k}$, and $\mathcal{B}_k$ represents the set of indices of variables in block $k$ $(k = 1, \ldots, K)$. The expansion coefficients have means $\mathbb{E}\{s_k\} = \mathbb{E}\{d_i\} = 0$, and variances $\mathbb{V}\{s_k\} = p_k \sigma_k^2 + \sigma^2$ and $\mathbb{V}\{d_i\} = O(\sigma^2)$, for $i = 1, \ldots, p - K$.

Note that if the conditions of the theorem are satisfied, then all treelets (*both* scaling and difference functions) associated with levels $\ell > p - K$ are constant on groups of similar variables. In particular, for a full decomposition at the maximum level $\ell = p - 1$ of the tree we have the following key result, which follows directly from Theorem 2:

COROLLARY 1. *Assume that the conditions in Theorem 2 are satisfied. A full treelet decomposition then gives $T^{p-1}(\Sigma) = s\phi + \sum_{i=1}^{p-1} d_i \psi_i$, where the scaling function $\phi$ and the $K - 1$ detail functions $\psi_{p-K+1}, \ldots, \psi_{p-1}$ are constant on each of the $K$ blocks. The coefficients $s$ and $d_{p-K+1}, \ldots, d_{p-1}$ reflect between-block structures, as opposed to the coefficients $d_1, \ldots, d_{p-K}$ which only reflect noise in the data with variances $\mathbb{V}\{d_i\} = O(\sigma^2)$ for $i = 1, \ldots, p - K$.*

The last result is interesting. It indicates a parameter-free way of finding $K$, the number of blocks, namely, by studying the *energy distribution* of a full treelet decomposition. Furthermore, the treelet transform can uncover the block structure even if it is hidden amidst a large number of background noise variables (see Figure 3 for a simulation with finite sample size).

REMARK 3. Both Theorem 2 and Corollary 1 can be directly generalized to include $p_0$ uncorrelated noise variables, so that $\mathbf{x} = (x_1, \ldots, x_{p-p_0}, x_{p-p_0+1}, \ldots, x_p)^T$, where $\mathbb{E}(x_i) = 0$ and $\mathbb{E}(x_i x_j) = 0$ for $i > p - p_0$ and $j \neq i$. For example, if equation (13) is satisfied, then the treelet decomposition at level $\ell = p - p_0$ is

$$T^{p-p_0}(\Sigma) = \sum_{k=1}^{K} s_k \phi_k + \sum_{i=1}^{p-p_0-K} d_i \psi_i + (0, \ldots, 0, x_{p-p_0+1}, \ldots, x_p)^T.$$

Furthermore, note that according to equation (41) in the Appendix A.3, within-block correlations are smallest ("worst-case scenario") when singletons are merged. Thus, the treelet transform is a *stabilizing* algorithm; once a few correct coarse-grained variables have been computed, it has the effect of denoising the data.

3.2.2. *Convergence rates.* The aim of this section is to give a rough estimate of the sample size required for treelets to discover the inherent structures of data. For covariance matrices with block structures, we show that treelets find the correct groupings of variables if the sample size $n \gg O(\log p)$, where $p$ is the dimension of the data. This is a significant result, as standard PCA—on the other hand—is consistent if and only if $p/n \to 0$ [Johnstone and Lu (2008)], that is, when $n \gg O(p)$. The result is also comparable to that in Bickel and Levina (2008) for regularization of sparse nearly diagonal covariance matrices. One main difference is that their paper assumes an a priori known ordered set of variables in which the covariance matrix is sparse, whereas treelets find such an ordering and coordinate system as part of the algorithm. The argument for treelets and a block covariance model goes as follows.

Assume that there are $K$ blocks in the population covariance matrix $\Sigma$. Define $A_{L,n}$ as the event that the $K$ maximum variance treelets, constructed at level $L = p - K$ of the tree, for a data set with $n$ observations, are supported *only* on variables from the same block. In other words, let $A_{L,n}$ represent the ideal case where the treelet transform finds the exact groupings of variables. Let $E_\ell$ denote the event that at level $\ell$ of the tree, the largest between-block sample correlation is less than the smallest within-block sample correlation,

$$E_\ell = \{\max \hat{\rho}_B^{(\ell)} < \min \hat{\rho}_W^{(\ell)}\}.$$

According to equations (31)–(32), the corresponding population correlations

$$\max \rho_B^{(\ell)} < \rho_1 \equiv \max_{1 \leq i,j \leq K} \left(\frac{\sigma_{ij}}{\sigma_i \sigma_j}\right), \qquad \min \rho_W^{(\ell)} > \rho_2 \equiv \frac{1}{\sqrt{1 + 3\max(\delta^2, \delta^4)}},$$

where $\delta = \frac{\sigma}{\min_k \sigma_k}$, for all $\ell$. Thus, a sufficient condition for $E_\ell$ is that $\{\max |\hat{\rho}_B^{(\ell)} - \rho_B^{(\ell)}| < t\} \cap \{\max |\hat{\rho}_W^{(\ell)} - \rho_W^{(\ell)}| < t\}$, where $t = (\rho_2 - \rho_1)/2 > 0$. We have that

$$\mathbb{P}(A_{L,n}) \geq \mathbb{P}\left(\bigcap_{0 \leq \ell < L} E_\ell\right)$$

$$\geq \mathbb{P}\left(\bigcap_{0 \leq \ell < L} \{\max |\hat{\rho}_B^{(\ell)} - \rho_B^{(\ell)}| < t\} \cap \{\max |\hat{\rho}_W^{(\ell)} - \rho_W^{(\ell)}| < t\}\right).$$

If (A1) holds, then it follows from Lemma 3 that

$$\mathbb{P}(A_{L,n}^C) \leq \sum_{0 \leq \ell < L} (\mathbb{P}(\max |\hat{\rho}_B^{(\ell)} - \rho_B^{(\ell)}| > t) + \mathbb{P}(\max |\hat{\rho}_W^{(\ell)} - \rho_W^{(\ell)}| > t))$$

$$\leq L c_1 p^2 e^{-n c_2 t^2}$$

for positive constants $c_1, c_2$. Thus, the requirement $\mathbb{P}(A_{L,n}^C) < \alpha$ is satisfied if the sample size

$$n \geq \frac{1}{c_2 t^2} \log\left(\frac{L c_1 p^2}{\alpha}\right).$$

From the large-sample properties of treelets (Section 3.1), it follows that treelets are consistent if $n \gg O(\log p)$.

**4. Treelets and a linear error-in-variables mixture model.** In this section we study a simple error-in-variables linear mixture model (factor model) which, under some conditions, gives rise to covariance matrices with block structures. Under this model, we compare treelets with PCA and variable selection methods. An advantage of introducing a concrete generative model is that we can easily relate our results to the underlying structures or components of real data; for example, different chemical substances in spectroscopy data, genes from the same pathway in microarray data, etc.

In light of this, consider a linear mixture model with $K$ components and additive noise. Each multivariate observation $\mathbf{x} \in \mathbb{R}^\mathbf{p}$ has the form

$$(15) \qquad \mathbf{x} = \sum_{j=1}^{K} u_j \mathbf{v}_j + \sigma \mathbf{z}.$$

The components or "factors" $u_j$ are random (but not necessarily independent) variables with variances $\sigma_j^2$. The "loading vectors" $\mathbf{v}_j$ are fixed, but typically unknown linearly independent vectors. In the last term, $\sigma$ represents the noise level, and $\mathbf{z} \sim \mathcal{N}_p(0, I)$ is a $p$-dimensional random vector.

In the *unsupervised* setting, we are given a training set $\{\mathbf{x}_i\}_{i=1}^n$ sampled from equation (15). Unsupervised learning tasks include, for example, inference on the number of components $K$, and on the underlying vectors $\mathbf{v}_j$. In the *supervised* setting, we consider a data set $\{\mathbf{x}_i, y_i\}_{i=1}^n$, where the response value $y$ of an observation $\mathbf{x}$ is a linear combination of the variables $u_j$ with a random noise term $\epsilon$,

$$(16) \qquad y = \sum_{j=1}^{K} \alpha_j u_j + \epsilon.$$

The standard supervised learning task in regression and classification is prediction of $y$ for new data $\mathbf{x}$, given a training set $\{\mathbf{x}_i, y_i\}_{i=1}^n$.

Linear mixture models are common in many fields, including spectroscopy and gene expression analysis. In spectroscopy equation (15) is known as Beer's law, where $\mathbf{x}$ is the logarithmic absorbance spectrum of a chemical substance measured at $p$ wavelengths, $u_j$ are the concentrations of constituents with pure absorbance spectra $\mathbf{v}_j$, and the response $y$ is typically one of the components, $y = u_i$. In gene data $\mathbf{x}$ is the measured expression level of $p$ genes, $u_j$ are intrinsic activities of various pathways, and each vector $\mathbf{v}_j$ represents the set of genes in a pathway. The quantity $y$ is typically some measure of severity of a disease, such as time until recurrence of cancer. A linear relation between $y$ and the values of $u_j$, as in equation (16), is commonly assumed.

4.1. *Treelets and a linear mixture model in the unsupervised setting.* Consider data $\{\mathbf{x}_i\}_{i=1}^n$ from the model in equation (15). Here we analyze an illustrative example with $K=3$ components and loading vectors $\mathbf{v}_k = I(\mathcal{B}_k)$, where $I$ is the indicator function, and $\mathcal{B}_k \subset \{1, 2, \ldots, p\}$ are sets of variables with sizes $p_k = |\mathcal{B}_k|$ ($k=1,2,3$). A more general analysis is possible but may not provide more insight.

The unsupervised task is to uncover the internal structure of the linear mixture model from data, for example, to infer the unknown structure of the vectors $\mathbf{v}_k$, including the sizes $p_k$ of the sets $\mathcal{B}_k$. The difficulty of this problem depends, among other things, on possible correlations between the random variables $u_j$, the variances of the components $u_j$, and interferences (overlap) between the loading vectors $\mathbf{v}_k$. We present three examples with increasing difficulty. Standard methods, such as principal component analysis, succeed only in the simplest case (Example 1), whereas more sophisticated methods, such as sparse PCA (elastic nets), sometimes require oracle information to correctly fit tuning parameters in the model. The treelet transform seems to perform well in all three cases. Moreover, the results are easy to explain by computing the covariance matrix of the data.

EXAMPLE 1 (Uncorrelated factors and nonoverlapping loading vectors). The simplest case is when the random variables $u_j$ are all uncorrelated for $j = 1, 2, 3$, and the loading vectors $\mathbf{v}_j$ are nonoverlapping. The population covariance matrix of $\mathbf{x}$ is then given by $\Sigma = C + \sigma^2 I_p$, where the noise-free matrix

$$(17) \qquad C = \begin{pmatrix} C_{11} & 0 & 0 & 0 \\ 0 & C_{22} & 0 & 0 \\ 0 & 0 & C_{33} & 0 \\ 0 & 0 & 0 & 0 \end{pmatrix}$$

is a $4 \times 4$ block matrix with the first three blocks $C_{kk} = \sigma_k^2 1_{p_k \times p_k}$ ($k = 1, 2, 3$), and the last diagonal block having all entries equal to zero.

Assume that $\sigma_k \gg \sigma$ for $k = 1, 2, 3$. This is a specific example of a *spiked covariance model* [Johnstone (2001)] the three components corresponding to distinct large eigenvalues or "spikes" of a model with background noise. As $n \to \infty$ with $p$ fixed, PCA recovers the hidden vectors $\mathbf{v}_1$, $\mathbf{v}_2$, and $\mathbf{v}_3$, since these three vectors exactly coincide with the principal eigenvectors of $\Sigma$. A treelet transform with a height $L$ determined by cross-validation and a normalized energy criterion returns the same results—which is consistent with Section 3.2 (Theorem 2 and Corollary 1).

The difference between PCA and treelets becomes obvious in the "small $n$, large $p$ regime." In the joint limit $p, n \to \infty$, standard PCA computes consistent estimators of the vectors $\mathbf{v}_j$ (in the presence of noise) if and only if $p(n)/n \to 0$ [Johnstone and Lu (2008)]. For an analysis of PCA for finite $p, n$, see, for example, Nadler (2007). As described in Section 3.2.2, treelets require asymptotically far fewer observations with the condition for consistency being $\log p(n)/n \to 0$.

EXAMPLE 2 (Correlated factors and nonoverlapping loading vectors). If the random variables $u_j$ are *correlated*, treelets are far better than PCA at inferring the underlying localized structure of the data—even asymptotically. Again, this is easy to explain and quantify by studying the data covariance structure. For example, assume that the loading vectors $\mathbf{v}_1$, $\mathbf{v}_2$, and $\mathbf{v}_3$ are nonoverlapping, but that the corresponding factors are dependent according to

$$(18) \qquad u_1 \sim N(0, \sigma_1^2), \qquad u_2 \sim N(0, \sigma_2^2), \qquad u_3 = c_1 u_1 + c_2 u_2.$$

The covariance matrix is then given by $\Sigma = C + \sigma^2 I_p$, where

$$(19) \qquad C = \begin{pmatrix} C_{11} & 0 & C_{13} & 0 \\ 0 & C_{22} & C_{23} & 0 \\ C_{13} & C_{23} & C_{33} & 0 \\ 0 & 0 & 0 & 0 \end{pmatrix}$$

with $C_{kk} = \sigma_k^2 1_{p_k \times p_k}$ (note that $\sigma_3^2 = c_1^2 \sigma_1^2 + c_2^2 \sigma_2^2$), $C_{13} = c_1 \sigma_1^2 1_{p_1 \times p_3}$, and $C_{23} = c_2 \sigma_2^2 1_{p_2 \times p_3}$. Due to the correlations between $u_j$, the loading vectors of the block model no longer coincide with the principal eigenvectors, and it is difficult to extract them with PCA.

We illustrate this problem by the example in Zou, Hastie and Tibshirani (2006). Specifically, let

$$(20) \qquad \begin{aligned} \mathbf{v}_1 &= [\overbrace{1\,1\,1\,1}^{\mathcal{B}_1} \ \overbrace{0\,0\,0\,0}^{\mathcal{B}_2} \ \overbrace{0\,0}^{\mathcal{B}_3}]^T, \\ \mathbf{v}_2 &= [0\,0\,0\,0 \ \ 1\,1\,1\,1 \ \ 0\,0]^T, \\ \mathbf{v}_3 &= [0\,0\,0\,0 \ \ 0\,0\,0\,0 \ \ 1\,1]^T, \end{aligned}$$

where there are $p = 10$ variables total, and the sets $\mathcal{B}_j$ are disjoint with $p_1 = p_2 = 4, p_3 = 2$ variables, respectively. Let $\sigma_1^2 = 290$, $\sigma_2^2 = 300$, $c_1 = -0.3$, $c_2 = 0.925$, and $\sigma = 1$. The corresponding variance $\sigma_3^2$ of $u_3$ is 282.8, and the covariances of the off-diagonal blocks are $\sigma_{13} = -87$ and $\sigma_{23} = 277.5$.

The first three PCA vectors for a training set of 1000 samples are shown in Figure 2 (left). It is difficult to infer the underlying vectors $\mathbf{v}_i$ from these results, as ideally, we would detect that, for example, the variables $(x_5, x_6, x_7, x_8)$ are all related and extract the latent vector $\mathbf{v}_2$ from only these variables. Simply thresholding the loadings and discarding small values also fails to achieve this goal [Zou, Hastie and Tibshirani (2006)]. The example illustrates the limitations of a global approach even with an infinite number of observations. In Zou, Hastie and Tibshirani (2006) the authors show by simulation that a combined $L_1$ and $L_2$-penalized least squares method, which they call sparse PCA or elastic nets, correctly identifies the sets of important variables if given "oracle information" on the number of variables $p_1, p_2, p_3$ in the different blocks. Treelets are similar in spirit to elastic nets as both methods tend to group highly correlated variables together. In this example the treelet algorithm is able to find both $K$, the number of components in the mixture model, and the hidden loading vectors $\mathbf{v}_i$—without any a priori knowledge or parameter tuning. Figure 2 (right) shows results from a treelet simulation with a large sample size ($n = 1000$) and a height $L = 7$ of the tree, determined by cross-validation (CV) and an energy criterion. The three maximum energy basis vectors correspond exactly to the hidden loading vectors in equation (20).

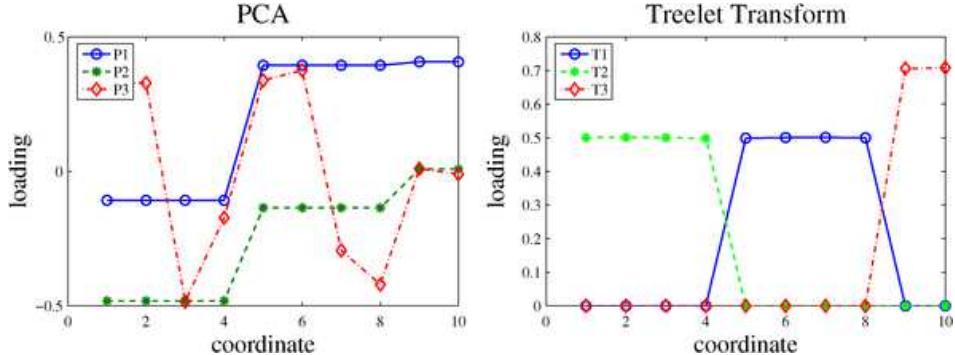

FIG. 2. *In Example 2 PCA fails to find the important variables in the three-component mixture model, as the computed eigenvectors* (left) *are sensitive to correlations between different components. On the other hand, the three maximum energy treelets* (right) *uncover the underlying data structures.*

EXAMPLE 3 (Uncorrelated factors and overlapping loading vectors). Finally, we study a challenging example where the first two loading vectors $\mathbf{v}_1$ and $\mathbf{v}_2$ are overlapping, the sample size $n$ is small, and the background noise level is high. Let $\{\mathcal{B}_1, \ldots, \mathcal{B}_4\}$ be disjoint subsets of $\{1, \ldots, p\}$, and let

(21) $\quad \mathbf{v}_1 = I(\mathcal{B}_1) + I(\mathcal{B}_2), \qquad \mathbf{v}_2 = I(\mathcal{B}_2) + I(\mathcal{B}_3), \qquad \mathbf{v}_3 = I(\mathcal{B}_4),$

where $I(\mathcal{B}_k)$ as before represents the indicator function for subset $k$ ($k = 1, \ldots, 4$). The population covariance matrix is then given by $\Sigma = C + \sigma^2 I_p$, where the noiseless matrix has the general form

(22) $$C = \begin{pmatrix} C_{11} & C_{12} & 0 & 0 & 0 \\ C_{12} & C_{22} & C_{23} & 0 & 0 \\ 0 & C_{23} & C_{33} & 0 & 0 \\ 0 & 0 & 0 & C_{44} & 0 \\ 0 & 0 & 0 & 0 & 0 \end{pmatrix},$$

with diagonal blocks $C_{11} = \sigma_1^2 1_{p_1 \times p_1}$, $C_{22} = (\sigma_1^2 + \sigma_2^2) 1_{p_2 \times p_2}$, $C_{33} = \sigma_2^2 1_{p_3 \times p_3}$, $C_{44} = \sigma_3^2 1_{p_4 \times p_4}$, and off-diagonal blocks $C_{12} = \sigma_1^2 1_{p_1 \times p_2}$ and $C_{23} = \sigma_2^2 1_{p_2 \times p_3}$. Consider a numerical example with $n = 100$ observations, $p = 500$ variables, and noise level $\sigma = 0.5$. We choose the same form for the components $u_1, u_2, u_3$ as in [Bair et al. (2006)], but associate the first two components with *overlapping* loading vectors $\mathbf{v}_1$ and $\mathbf{v}_2$. Specifically, the components are given by $u_1 = \pm 0.5$ with equal probability, $u_2 = I(U_2 < 0.4)$, and $u_3 = I(U_3 < 0.3)$, where $I(x)$ is the indicator of $x$, and $U_j$ are all independent uniform random variables in [0,1]. The corresponding variances are $\sigma_1^2 = 0.25$, $\sigma_2^2 = 0.24$, and $\sigma_3^2 = 0.21$. As for the blocks $\mathcal{B}_k$, we consider $\mathcal{B}_1 = \{1, \ldots, 10\}, \mathcal{B}_2 = \{11, \ldots, 50\}, \mathcal{B}_3 = \{51, \ldots, 100\}$, and $\mathcal{B}_4 = \{201, \ldots, 400\}$.

Inference in this case is challenging for several different reasons. The sample size $n < p$, the loading vectors $\mathbf{v}_1$ and $\mathbf{v}_2$ are overlapping in the region $\mathcal{B}_2 = \{11, \ldots, 50\}$, and the signal-to-noise ratio is low with the variance $\sigma^2$ of the noise essentially being of the same size as the variances $\sigma_j^2$ of $u_j$. Furthermore, the condition in equation (13) is not satisfied even for the population covariance matrix. Despite these difficulties, the treelet algorithm is remarkably stable, returning results that by and large correctly identify the internal structures of the data. The details are summarized below.

Figure 3 (top center) shows the energy score of the best $K$-basis at different levels of the tree. We used 5-fold cross-validation; that is, we generated a single data set of $n = 100$ observations, but in each of the 5 computations the treelets were constructed on a subset of 80 observations, with 20 observations left out for the energy score computation. The five curves as well as their average clearly indicate a "knee" at the level $L = 300$. This is consistent with our expectations that the treelet algorithm mainly merges noise variables at levels $L \geq |\bigcup_k \mathcal{B}_k|$. For a tree with "optimum" height $L = 300$, as indicated by the CV results, we then constructed a treelet basis on the full data set. Figure 3 (top right) shows the energy of these treelets sorted according to descending energy score. The results indicate that we have two dominant treelets, while the remaining treelets have an energy that is either slightly higher or of the same order as the variance of the noise. In Figure 3 (bottom left) we plot the loadings of the four highest energy treelets. "Treelet 1" (red) is approximately constant on the set $\mathcal{B}_4$ (the support of $\mathbf{v}_3$), "Treelet 2" (blue) is approximately piecewise constant on blocks $\mathcal{B}_1$, $\mathcal{B}_2$, and $\mathcal{B}_3$ (the support of $\mathbf{v}_1$ and $\mathbf{v}_2$), while the low-energy degenerate treelets 3 (green) and 4 (magenta) seem to take differences between variables in the sets $\mathcal{B}_1$, $\mathcal{B}_2$, and $\mathcal{B}_3$. Finally, we computed 95% confidence bands of the treelets using 1000 bootstrap samples and the method described in Section 3.1. Figure 3 (bottom right) indicate, that the treelet results for the two maximum energy treelets are rather stable despite the small sample size and the low signal-to-noise ratio. Most of the time the first treelet selects variables from $\mathcal{B}_4$, and most of the time the second treelet selects variables from $\mathcal{B}_2$ and either $\mathcal{B}_1$ or $\mathcal{B}_3$ or both sets. The low-energy treelets seem to pick up differences between blocks $\mathcal{B}_1$, $\mathcal{B}_2$, and $\mathcal{B}_3$, but the exact order in which they select the variables vary from simulation to simulation. As described in the next section, for the purpose of regression, the main point is that the *linear span* of the first few highest energy treelets is a good approximation of the span of the unknown loading vectors, $\text{Span}\{\mathbf{v}_1, \ldots, \mathbf{v}_K\}$.

4.2. *The treelet transform as a feature selection scheme prior to regression.* Knowing some of the basic properties of treelets, we now examine a typical regression or classification problem with data $\{\mathbf{x}_i, y_i\}_{i=1}^n$ given by equations (15) and (16). As the data $\mathbf{x}$ are noisy, this is an error-in-variables type problem. Given a training set, the goal is to construct a linear function $f : \mathbb{R}^p \to \mathbb{R}$ to predict $\hat{y} = f(\mathbf{x}) = \mathbf{r} \cdot \mathbf{x} + b$ for a new observation $\mathbf{x}$.

Before considering the performance of treelets and other algorithms in this setting, we review some of the properties of the optimal mean-squared error (MSE) predictor. For simplicity, we consider the case $y = u_1$ in equation (16), and denote by $P_1 : \mathbb{R}^p \to \mathbb{R}^p$ the projection operator onto the space spanned by the vectors $\{\mathbf{v}_2, \ldots, \mathbf{v}_K\}$. In this setting the unbiased MSE-optimal estimator has a regression vector $\mathbf{r} = \mathbf{v}_y / \|\mathbf{v}_y\|^2$, where $\mathbf{v}_y = \mathbf{v}_1 - P_1 \mathbf{v}_1$. The vector $\mathbf{v}_y$ is the part of the loading vector $\mathbf{v}_1$ that is unique to the response variable $y = u_1$, since the projection of $\mathbf{v}_1$ onto the span of the loading vectors of the other components $(u_2, \ldots, u_K)$ has been subtracted. For example, in the case of only two components, we have that

$$\mathbf{v}_y = \mathbf{v}_1 - \frac{\mathbf{v}_1 \cdot \mathbf{v}_2}{\|\mathbf{v}_2\|^2} \mathbf{v}_2. \tag{23}$$

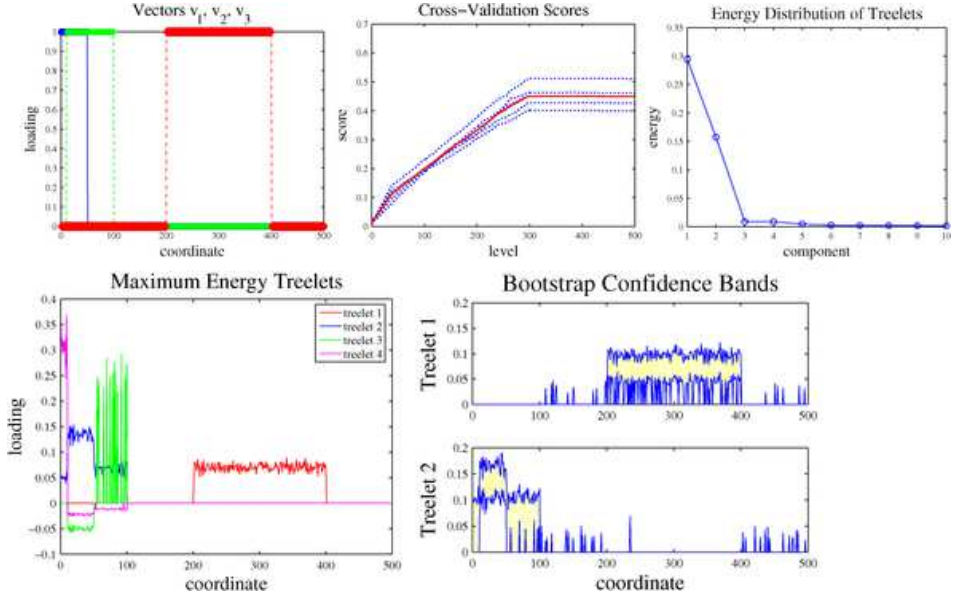

FIG. 3. Top left: *The vectors $\mathbf{v}_1$ (blue), $\mathbf{v}_2$ (green), $\mathbf{v}_3$ (red) in Example 3.* Top center: *The "score" or total energy of $K = 3$ maximum variance treelets computed at different levels of the tree with 5-fold cross-validation; dotted lines represent the five different simulations and the solid line the average score.* Top right: *Energy distribution of the treelet basis for the full data set at an "optimal" height $L = 300$.* Bottom left: *The four treelets with highest energy.* Bottom right: *95% confidence bands by bootstrap for the two dominant treelets.*

The vector $\mathbf{v}_y$ plays a central role in chemometrics, where it is known as the *net analyte signal* [Lorber, Faber and Kowalski (1997), Nadler and Coifman (2005a)]. Using this vector for regression yields a mean squared error of prediction

$$\mathbb{E}\{(\hat{y} - y)^2\} = \frac{\sigma^2}{\|\mathbf{v}_y\|^2}. \tag{24}$$

We remark that, similar to shrinkage in point estimation, there exist biased estimators with smaller MSE [Gruber (1998), Nadler and Coifman (2005b)], but for large signal to noise ratios $(\sigma/\|\mathbf{v}_y\| \ll 1)$, such shrinkage is negligible.

Many regression methods [including multivariate least squares, partial least squares (PLS), principal component regression (PCR), etc.] attempt to compute the optimal regression vector or net analyte signal (NAS). It can be shown that in the limit $n \to \infty$, both PLS and PCR are MSE-optimal. However, in some applications, the number of variables is much larger than the number of observations ($p \gg n$). The question at hand is then, what the effect of small sample size is on these methods, when combined with noisy high-dimensional data. Both PLS and PCR first perform a global dimensionality reduction from $p$ to $k$ variables, and then apply least squares linear regression on these $k$ features. As described in Nadler and Coifman (2005b), their main limitation is that in the presence of noisy high dimensional data, the computed projections are noisy themselves. For fixed $p$ and $n$, a Taylor expansion of the regression coefficient as a function of the noise level $\sigma$ shows that these methods have an averaged prediction error

$$\mathbb{E}\{(\hat{y} - y)^2\} \simeq \frac{\sigma^2}{\|\mathbf{v}_y\|^2} \left[ 1 + \frac{c_1}{n} + \frac{c_2 \sigma^2}{\mu \|\mathbf{v}_y\|^2} \frac{p^2}{n^2} (1 + o(1)) \right]. \tag{25}$$

In equation (25) the coefficients $c_1$ and $c_2$ are both $O(1)$ constants, independent of $\sigma$, $p$, and $n$. The quantity $\mu$ depends on the specific algorithm used, and is a measure of the variances and covariances of the different components $u_j$, and of the amount of interferences of their loading vectors $\mathbf{v}_j$. The key point of this analysis is that when $p \gg n$, the last term in (25) can dominate and lead to large prediction errors. This emphasizes the limitations of global dimensionality reduction methods, and the need for robust feature selection and dimensionality reduction of the data *prior* to application of learning algorithms such as PCR and PLS.

Other common approaches to dimensionality reduction in this setting are *variable selection schemes*, specifically those that choose a small subset of variables based on their individual correlation with the response $y$. To analyze their performance, we consider a more general dimensionality reduction transformation $T : \mathbb{R}^p \to \mathbb{R}^k$ defined by $k$ orthonormal projections $\mathbf{w}_i \in \mathbb{R}^p$,

$$(26) \qquad T\mathbf{x} = (\mathbf{x} \cdot \mathbf{w}_1, \mathbf{x} \cdot \mathbf{w}_2, \ldots, \mathbf{x} \cdot \mathbf{w}_k).$$

This family of transformations includes variable subset selection methods, where each projection $\mathbf{w}_j$ selects one of the original variables. It also includes wavelet methods and our proposed treelet transform. Since an orthonormal projection of a Gaussian noise vector in $\mathbb{R}^p$ is a Gaussian vector in $\mathbb{R}^k$, and a relation similar to equation (15) holds between $T\mathbf{x}$ and $y$, formula (25) still holds, but with the original dimension $p$ replaced by $k$, and with $\mathbf{v}_y$ replaced by its projection $T\mathbf{v}_y$,

$$(27) \qquad \mathbb{E}\{(\hat{y} - y)^2\} \simeq \frac{\sigma^2}{\|T\mathbf{v}_y\|^2}\left[1 + \frac{c_1}{n} + \frac{c_2\,\sigma^2}{\mu\|T\mathbf{v}_y\|^2}\frac{k^2}{n^2}(1 + o(1))\right].$$

Equation (27) indicates that a dimensionality reduction scheme should ideally preserve the net analyte signal of $y$ ($\|T\mathbf{v}_y\| \simeq \|\mathbf{v}_y\|$), while at the same time represent the data by as few features as possible ($k \ll p$).

The main problem of PCA is that it optimally fits the noisy data, yielding for the noise-free response $\|T\mathbf{v}_y\|/\|\mathbf{v}_y\| \simeq (1 - c\sigma^2 p^2/n^2)$. The main limitation of variable subset selection schemes is that in complex settings with overlapping vectors $\mathbf{v}_j$, such schemes may at best yield $\|T\mathbf{v}_y\|/\|\mathbf{v}_y\| < 1$. Due to high dimensionality, the latter methods may still achieve better prediction errors than methods that use all the original variables. However, with a more general variable transformation/compression method, one could potentially better capture the NAS. If the data $\mathbf{x}$ are a priori known to be smooth continuous signals, a reasonable choice is wavelet compression, which is known to be asymptotically optimal. In the case of unstructured data, we propose to use treelets.

To illustrate these points, we revisit Example 3 in Section 4.1, and compare treelets to the variable subset selection scheme of Bair et al. (2006) for PLS, as well as global PLS on all variables. As before, we consider a relatively small training set of size $n = 100$ but here we include 1500 additional noise variables, so that $p = 2000 \gg n$. We furthermore assume that the response is given by $y = 2u_1$. The vectors $\mathbf{v}_j$ are shown in Figure 3 (top left). The two vectors $\mathbf{v}_1$ and $\mathbf{v}_2$ overlap, but $\mathbf{v}_1$ (associated with the response) and $\mathbf{v}_3$ are orthogonal. Therefore, the response vector unique to $y$ (the net analyte signal) is given by equation (23); see Figure 4 (left).

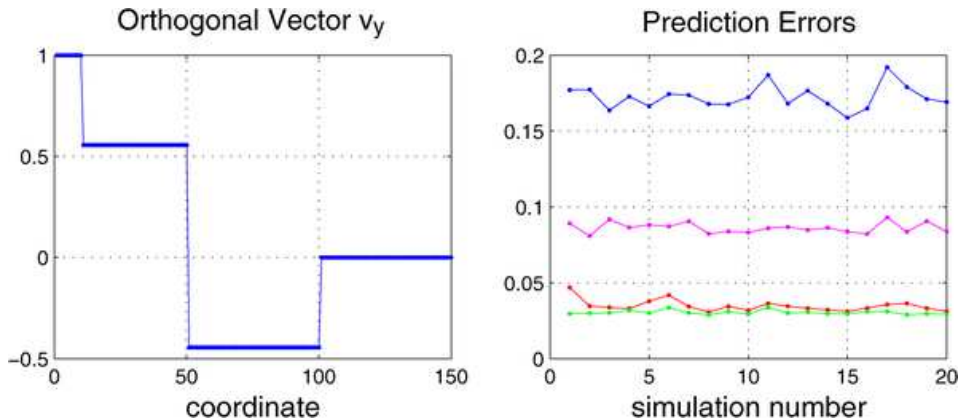

FIG. 4. *Left: The vector $\mathbf{v}_y$ (only the first 150 coordinates are shown as the rest are zero). Right: Averaged prediction errors of 20 simulation results for the methods, from top to bottom: PLS on all variables (blue), supervised PLS with variable selection (purple), PLS on treelet features (green), and PLS on projections onto the true vectors $\mathbf{v}_i$ (red).*

To compute $\mathbf{v}_y$, all the 100 first coordinates (the set $\mathcal{B}_1 \cup \mathcal{B}_2 \cup \mathcal{B}_3$) are needed. However, a feature selection scheme that chooses variables based on their correlation to the response will pick the first 10 coordinates and then the next 40, that is, *only* variables in the set $\mathcal{B}_1 \cup \mathcal{B}_2$ (the support of the loading vector $\mathbf{v}_1$). Variables numbered 51 to 100 (set $\mathcal{B}_3$), although critical for prediction of the response $y = 2u_1$, are uncorrelated with it (as $u_1$ and $u_2$ are uncorrelated) and are thus not chosen, even in the limit $n \to \infty$. In contrast, even in the presence of moderate noise and a relatively small sample size of $n = 100$, the treelet algorithm correctly joins together the subsets of variables 1–10, 11–50, 51–100 and 201–400 (i.e., variables in the sets $\mathcal{B}_1, \mathcal{B}_2, \mathcal{B}_3, \mathcal{B}_4$). The rest of the variables, which contain only noise, are combined only at much higher levels in the treelet algorithm, as they are asymptotically uncorrelated. Because of this, using only coarse-grained sum variables in the treelet transform yields near optimal prediction errors. In Figure 4 (right) we plot the mean squared error of prediction (MSEP) for 20 different simulations with prediction error computed on an independent test set of 500 observations. The different methods are PLS on all variables (MSEP = 0.17), supervised PLS with variable selection as in Bair et al. (2006) (MSEP = 0.09), PLS on the 50 treelet features with highest variance, with the level of the treelet determined by leave-one-out cross validation (MSEP = 0.035), and finally PLS on the projection of the noisy data onto the true vectors $\mathbf{v}_i$, assuming they were known (MSEP = 0.030). In all cases, the optimal number of PLS projections (latent variables) is also determined by leave-one-out cross validation. Due to the high dimensionality of the data, choosing a subset of the original variables performs better than full-variable methods. However, choosing a subset of *treelet features* performs even better, yielding an almost optimal prediction error $(\sigma^2/\|\mathbf{v}_y\|^2 \approx 0.03)$; compare the green and red curves in the figure.

## 5. Examples.

5.1. *Hyperspectral analysis and classification of biomedical tissue.* To illustrate how our method works for data with highly complex dependencies between variables, we use an example from hyperspectral imaging of biomedical tissue. Here we analyze a hyperspectral image of an H&E stained microarray section of normal human colon tissue [see Angeletti et al. (2005) for details on the data collection method]. This is an ordered data set of moderate to high dimension. One scan of the tissue specimen returns a $1024 \times 1280$ data cube or "hyperspectral image," where each pixel location contains spectral measurements at 28 known frequencies between 420 nm and 690 nm. These spectra give information about the chemical structure of the tissue. There is, however, redundancy as well as noise in the spectra. The challenge is to find the right coordinate system for this relatively high-dimensional space, and extract coordinates (features) that contain the most useful information about the chemicals and substances of interest.

We consider the problem of tissue discrimination using only spectral information. With the help of a pathologist, we manually label about 60000 pixels of the image as belonging to three different tissue types (colon cell nuclei, cytoplasm of colon cells, cytoplasm of goblet cells). Figure 5 shows the locations of the labeled pixels, and their tissue-specific transmission spectra. Figure 6 shows an example of how treelets can learn the covariance structure for colon cell nuclei (Tissue type 3). The method learns both the tree structure and a basis through a series of Jacobi rotations (see top right panel). By construction, the basis vectors are *localized* and supported on nested clusters in the tree (see the bottom left and top left panels). As a comparison, we have also computed the PCA eigenvectors. The latter vectors are global and involve all the original variables (see bottom right panel).

In a similar way, we apply the treelet transform to the training data in a 5-fold cross-validation test on the full data set with labeled spectra: Using a (maximum height) treelet decomposition, we construct a basis for the training set in each fold. To each basis vector, we assign a *discriminant score* that quantifies how well it distinguishes spectra from two different tissue types. The total score for vector $\mathbf{w}_i$ is defined as

$$\hat{\mathcal{E}}(\boldsymbol{w}_i) = \sum_{j=1}^{K} \sum_{k=1; k \neq j}^{K} H(\hat{p}_i^{(j)} \| \hat{p}_i^{(k)}), \tag{28}$$

where $K = 3$ is the number of classes, and $H(\hat{p}_i^{(j)} \| \hat{p}_i^{(k)})$ is the Kullback–Leibler distance between the estimated marginal density functions $\hat{p}_i^{(j)}$ and $\hat{p}_i^{(k)}$ of class-j and class-k signals, respectively, in the direction of $\boldsymbol{w}_i$. We project our training data onto the $K$ ($< 28$) most discriminant directions, and build a Gaussian classifier in this reduced feature space. This classifier is finally used to label the test data and to estimate the misclassification error rate. The left panel in Figure 7 shows the average CV error rate as a function of the number of local discriminant features. (As a comparison, we show similar results for Haar–Walsh wavelet packets and a local discriminant basis [Saito, Coifman, Geshwind and Warner (2002)] which use the same discriminant score to search through a library of orthonormal wavelet bases.) The straight line represents the error rate if we apply a Gaussian classifier directly to the 28 components in the original coordinate system. The key point is that, with 3 treelet features, we get the same performance as if we used all the original data. Using more treelet features yields an *even* lower misclassification rate. (Because of the large sample size, the curse of dimensionality is not noticeable for $< 15$ features.) These results indicate that a treelet representation has advantages beyond the obvious benefits of a dimensionality reduction. We are effectively "denoising" the data by changing our coordinate system and discarding irrelevant coordinates. The right panel in Figure 7 shows the three most discriminant treelet vectors for the full data set. These vectors resemble continuous-valued versions of the indicator functions in Section 3.2. Projecting onto one of these vectors has the effect of first taking a weighted average of adjacent spectral bands, and then computing a difference between averages of bands in different regions of the spectrum. (In Section 5.3, Figure 10, we will see another example that the loadings themselves contain information about structure in data.)

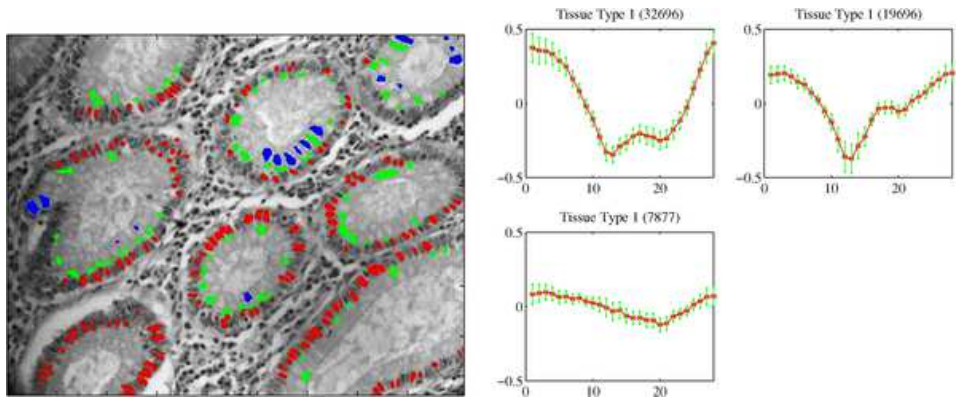

FIG. 5. Left: *Microscopic image of a cross-section of colon tissue. At each pixel position, the spectral characteristics of the tissue is measured at 28 different wavelengths ($\lambda = 420, 430, \ldots, 690$ nm). For our analysis, we manually label about* 60000 *individual spectra: Red marks the locations of spectra of "Tissue type 1" (nuclei), green "Tissue type 2" (cytoplasm of colon cells), and blue corresponds to samples of "Tissue type 3" (cytoplasm of goblet cells).* Right: *Spectral signatures of the 3 different tissue types. Each plot shows the sample mean and standard deviation of the log-transmission spectra.*

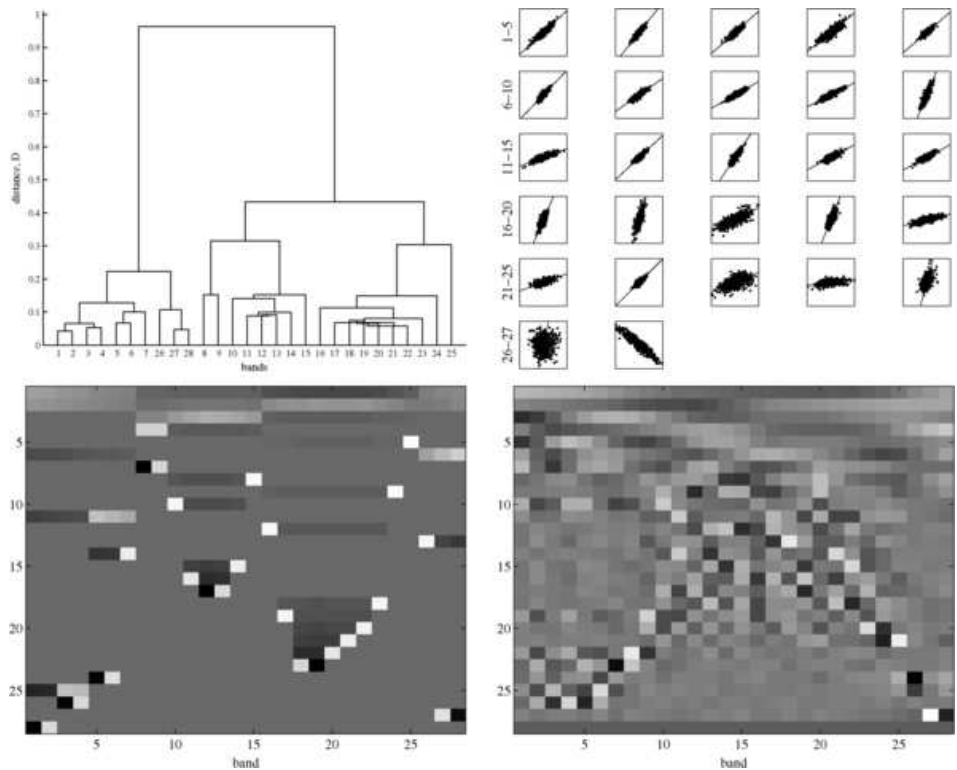

FIG. 6. Top left: *Learned tree structure for nuclei (Tissue Type 1). In the dendrogram the height of each U-shaped line represents the distance* $d_{ij} = (1 - \rho_{ij})/2$, *where* $\rho_{ij}$ *is the correlation coefficient for the two variables combined. The leaf nodes represent the* $p = 28$ *original spectral bands.* Top right: *2D scatter plots of the data at levels* $\ell = 1, \ldots, p-1$. *Each plot shows 500 randomly chosen data points; the lines indicate the first principal directions and rotations relative to the variables that are combined. (Note that a Haar wavelet corresponds to a fixed* $\pi/4$ *rotation.)* Bottom left: *Learned orthonormal basis. Each row represents a localized vector, supported on a cluster in the hierarchical tree.* Bottom right: *Basis computed by a global eigenvector analysis (PCA).*

5.2. *A classification example with an internet advertisement data set.* Here we study an internet advertisement data set from the UCI ML repository [Kushmerick (1999)]. This is an example of an unordered data set of high dimension where many variables are collinear. After removal of the first three continuous variables, this set contains 1555 binary variables and 3279 observations, labeled as belonging to one of two classes. The goal is to predict whether a new observation (an image in an internet page) is an internet

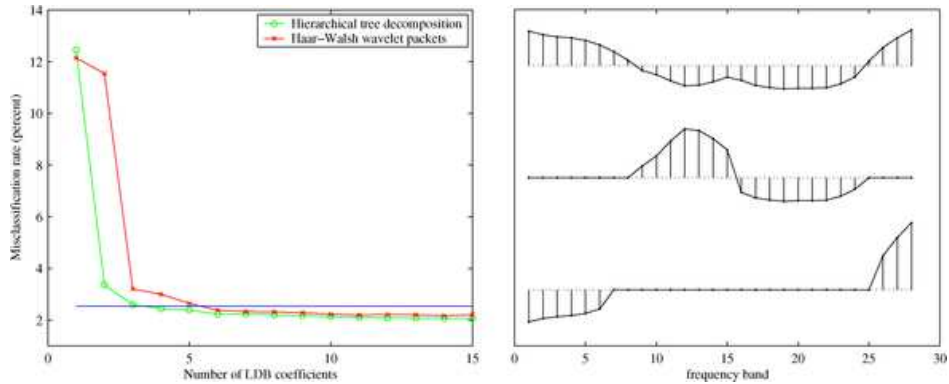

FIG. 7. Left: *Average misclassification rate (in a 5-fold cross-validation test) as a function of the number of top discriminant features retained, for a treelet decomposition (rings), and for Haar-Walsh wavelet packets (crosses). The constant level around 2.5% indicates the performance of a classifier directly applied to the 28 components in the original coordinate system.* Right: *The top 3 local discriminant basis (LDB) vectors in a treelet decomposition of the full data set.*

TABLE 1
*Classification test errors for an internet advertisement data set*

| Classifier | Full data set (1555 variables) | Reduced data set (760 variables) | Final representation with coarse-grained treelet features |
|---|---|---|---|
| LDA | 5.5% | 5.1% | 4.5% |
| 1-NN | 4.0% | 4.0% | 3.7% |

advertisement or not, given values of its 1555 variables (various features of the image).

With standard classification algorithms, one can easily obtain a generalization error of about 5%. The first column in Table 1, labeled "full data set," shows the misclassification rate for linear discriminant analysis (LDA) (with the additional assumption of a diagonal covariance matrix), and for 1-nearest neighbor (1-NN) classification. The average is taken over 25 randomly selected training and test sets, with 3100 and 179 observations each.

The internet-ad data set has several distinctive properties that are clearly revealed by an analysis with treelets: First of all, several of the original variables are exactly linearly related. As the data are binary ($-1$ or 1), these variables are either identical or of opposite values. In fact, one can reduce the dimensionality of the data from 1555 to 760 without loss of information. The second column in the table labeled "reduced data set" shows the decrease in error rate after a lossless compression where we have simply removed redundant variables. Furthermore, of these remaining 760 variables, many are highly related, with subsets of similar variables. The treelet algorithm automatically identifies these groups, as the algorithm reorders the variables during the basis computation, encoding the information in such a group with a coarse-grained sum variable and difference variables for the residuals. Figure 8, left, shows the correlation matrix of the first 200 out of 760 variables in the order they are given. To the right, we see the corresponding matrix, after sorting all variables according to the order in which they are combined by the treelet algorithm. Note how the (previously hidden) block structures "pop out."

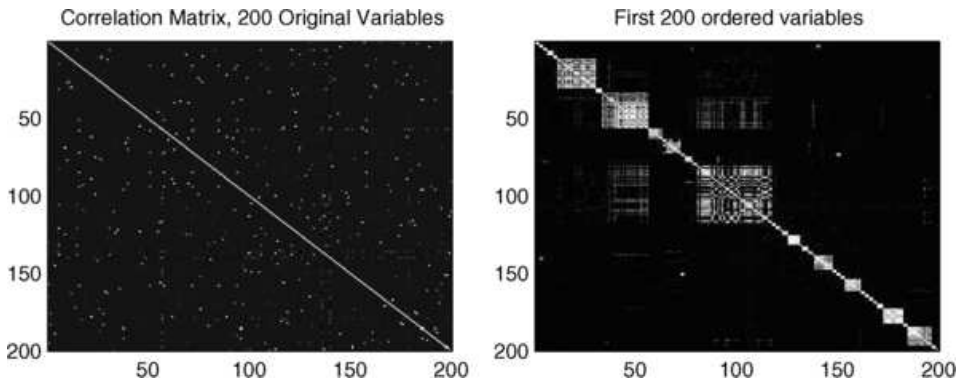

Fig. 8. Left: *The correlation matrix of the first 200 out of 760 variables in the order they were originally given.* Right: *The corresponding matrix, after sorting all variables according to the order in which they are combined by the treelet algorithm.*

A more detailed analysis of the reduced data set with 760 variables shows that there are more than 200 distinct pairs of variables with a correlation coefficient larger than 0.95. Not surprisingly, as shown in the right column of Table 1, treelets can further increase the predictive performance on this data set, yielding results competitive with other feature selection methods in the literature [Zhao and Liu (2007)]. All results in Table 1 are averaged over 25 different simulations. As in Section 4.2, the results are achieved at a level $L < p-1$, by projecting the data onto the treelet scaling functions, that is, by only using coarse-grained sum variables. The height $L$ of the tree is found by 10-fold cross-validation and a minimum prediction error criterion.

5.3. *Classification and analysis of DNA microarray data.* We conclude with an application to DNA microarray data. In the analysis of gene expression, many methods first identify groups of highly correlated variables and then choose a few representative genes for each group (a so-called gene signature). The treelet method also identifies subsets of genes that exhibit similar expression patterns, but in contrast, replaces each such localized group by a linear combination that encodes the information from all variables in that group. As illustrated in previous examples in the paper, such a representation typically regularizes the data which improves the performance of regression and classification algorithms.

Another advantage is that the treelet method yields a *multi-scale* data representation well-suited for the application. The benefits of hierarchical clustering in exploring and visualizing microarray data are well recognized in the field [Eisen et al. (1998), Tibshirani et al. (1999)]. It is, for example, known that a hierarchical clustering (or dendrogram) of genes can sometimes reveal interesting clusters of genes worth further investigation. Similarly, a dendrogram of samples may identify cases with similar medical conditions. The treelet algorithm automatically yields such a re-arrangement and interpretation of the data. It also provides an orthogonal basis for data representation and compression.

We illustrate our method on the leukemia data set of Golub et al. (1999). This data monitor expression levels for 7129 genes and 72 patients, suffering from acute lymphoblastic leukemia (ALL, 47 cases) or acute myeloid leukemia (AML, 25 cases). The data are known to have a low intrinsic dimensionality, with groups of genes having similar expression patterns across samples (cell lines). The full data set is available at http://www.genome.wi.mit.edu/MPR, and includes a training set of 38 samples and a test set of 34 samples.

Prior to analysis, we use a standard two-sample $t$-test to select genes that are differentially expressed in the two leukemia types. Using the training data, we perform a full (i.e., maximum height) treelet decomposition of the $p = 1000$ most "significant" genes. We sort the treelets according to their energy content [equation (5)] on the training samples, and project the test data onto the $K$ treelets with the highest energy score. The reduced data representation of each sample (from $p$ genes to $K$ features) is finally used to classify the samples into the two leukemia types, ALL or AML. We examine two different classification schemes:

In the first case, we apply a linear Gaussian classifier (LDA). As in Section 5.2, the treelet transform serves as a feature extraction and dimensionality reduction tool prior to classification. The appropriate value of the dimension $K$ is chosen by 10-fold cross-validation (CV). We divide the training set at random into 10 approximately equal-size parts, perform a separate $t$-test in each fold, and choose the $K$-value that leads to the smallest CV classification error (Figure 9, left).

In the second case, we classify the data using a novel *two-way treelet decomposition scheme*: we first compute treelets on the genes, then we compute treelets on the samples. As before, each sample (patient) is represented by $K$ treelet features instead of the $p$ original genes. The dimension $K$ is chosen by cross-validation on the training set. However, instead of applying a standard classifier, we construct treelets *on the samples* using the new patient profiles. The two main branches of the associated dendrogram divide the samples into two classes, which are labeled using the training data and a majority vote. Such a two-way decomposition—of both genes and samples—leads to classification results competitive with other algorithms; see Figure 9, right, and Table 2 for a comparison with benchmark results in

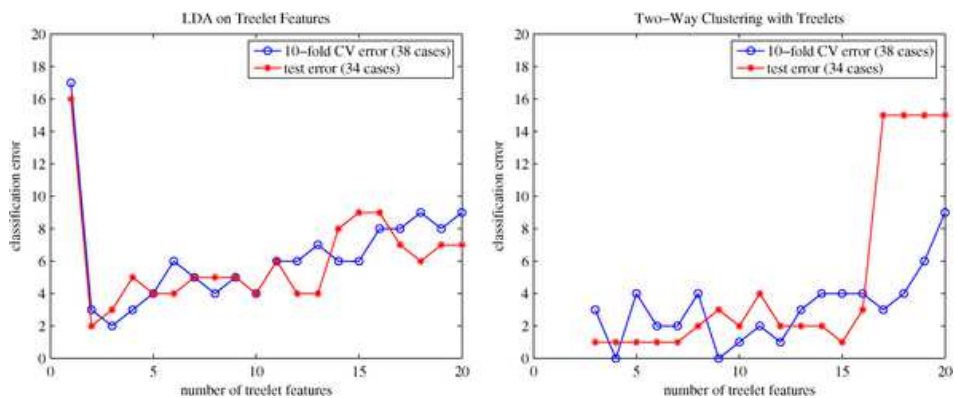

FIG. 9. *Number of misclassified cases as a function of the number of treelet features.* Left: *LDA on treelet features; ten-fold cross-validation gives the lowest misclassification rate (2/38) for $K=3$ treelets; the test error rate is then 3/34.* Right: *Two-way decomposition of both genes and samples; the lowest CV misclassification rate (0/38) is for $K=4$; the test error rate is then 1/34.*

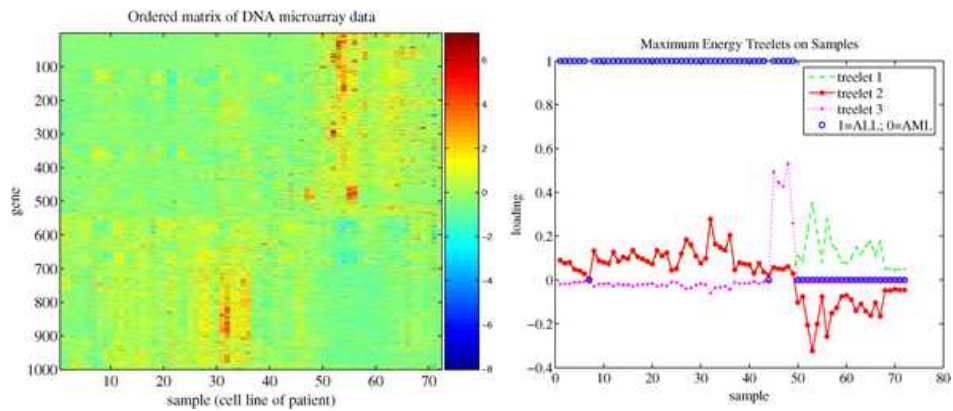

FIG. 10. Left, *the gene expression data with rows (genes) and columns (samples) ordered according to a hierarchical two-way clustering with treelets. (For display purposes, the expression levels for each gene are here normalized across the samples to zero mean and unit standard deviation.)* Right, *the three maximum energy treelets on ordered samples. The loadings of the highest-energy treelet (red) is a good predictor of the true labels (blue circles).*

Table 2
*Leukemia misclassification rates; courtesy of Zou and Hastie (2005)*

| Method | Ten-fold CV error | Test error |
| --- | --- | --- |
| Golub et al. (1999) | 3/38 | 4/34 |
| Support vector machines (Guyon et al. (2002)) | 2/38 | 1/34 |
| Nearest shrunken centroids (Tibshirani et al. (2002)) | 2/38 | 2/34 |
| Penalized logistic regression (Zhu and Hastie (2004)) | 2/38 | 1/34 |
| Elastic nets (Zou and Hastie (2005)) | 3/38 | 0/34 |
| LDA on treelet features | 2/38 | 3/34 |
| Two-way treelet decomposition | 0/38 | 1/34 |

Zou and Hastie (2005). Moreover, the proposed method returns orthogonal functions with continuous-valued information on hierarchical groupings of genes or samples.

Figure 10 (left) displays the original microarray data, with rows (genes) and columns (samples) ordered according to a hierarchical two-way clustering with treelets. The graph to the right shows the three maximum energy treelets on ordered samples. Note that the loadings are small for the two cases that are misclassified. In particular, "Treelet 2" is a good "continuous-valued" indicator function of the true classes. The results for the treelets on genes are similar. The key point is that whenever there is a group of highly correlated variables (genes or samples), the algorithm tends to choose a coarse-grained variable for that whole group (see, e.g., "Treelet 3" in the figure). The weighting is *adaptive*, with loadings that reflect the complex internal data structure.

**6. Conclusions.** In the paper we described a variety of situations where the treelet transform outperforms PCA and some common variable selection methods. The method is especially useful as a feature extraction and regularization method in situations where variables are collinear and/or the data is noisy with the number of variables, $p$, far exceeding the number of observations, $n$. The algorithm is fully adaptive, and returns both a hierarchical tree and loading functions that reflect the internal localized structure of the data. We showed that, for a covariance model with block structure, the maximum energy treelets converge to a solution where they are constant on each set of indistinguishable variables. Furthermore, the convergence rate of treelets is considerably faster than PCA, with the required sample size for consistency being $n \gg O(\log p)$ instead of $n \gg O(p)$. Finally, we demonstrated the applicability of treelets on several real data sets with highly complex dependencies of variables.

# APPENDIX

**A.1. Proof of Theorem 1.** Let $\mathbf{x} = (x_1, \ldots, x_p)^T$ be a random vector with distribution $F$ and covariance matrix $\Sigma = \Sigma_F$. Let $\rho_{ij}$ denote the correlation between $x_i$ and $x_j$. Let $\mathbf{x}^1, \ldots, \mathbf{x}^n$ be a sample from $F$, and denote the sample covariance matrix and sample correlations by $\hat{\Sigma}$ and $\hat{\rho}_{ij}$. Let $\mathcal{S}_p$ denote all $p \times p$ covariance matrices. Let

$$\mathcal{F}_n(b) = \left\{ F : \Sigma_F \text{ is positive definite, } \min_{1 \le j \le p_n} \sigma_j \ge b \right\}.$$

Any of the assumptions (A1a), (A1b), or (A1c) are sufficient to guarantee certain exponential inequalities.

LEMMA A.1. *There exist positive constants $c_1, c_2$ such that, for every $\epsilon > 0$,*

$$\mathbb{P}(\|\hat{\Sigma}_{jk} - \Sigma_{jk}\|_\infty > \epsilon) \le c_1 p_n^2 e^{-nc_2 \epsilon^2}. \tag{29}$$

*Hence,*

$$\|\hat{\Sigma}_{jk} - \Sigma_{jk}\|_\infty = O_P\left(\sqrt{\frac{\log n}{n}}\right).$$

PROOF. Under (A1), (29) is an immediate consequence of standard exponential inequalities and the union bound. The last statement follows by setting $\epsilon_n = K\sqrt{\log n / n}$ for sufficiently large $K$ and applying (A2). □

LEMMA A.2. *Assume either that* (i) *$x$ is multivariate normal or that* (ii) *$\max_{1 \le j \le p} |x_j| \le B$ for some finite $B$ and $\min_j \sigma_j \ge b > 0$. Then, there exist positive constants $c_3, c_4$ such that, for every $\epsilon > 0$,*

$$\mathbb{P}\left(\max_{jk} |\hat{\rho}_{jk} - \rho_{jk}| > \epsilon\right) \le c_3 p^2 e^{-nc_4 \epsilon^2}. \tag{30}$$

PROOF. Under normality, this follows from Kalisch and Bühlmann (2007). Under (ii) note that $h(\sigma_1, \sigma_2, \sigma_{12}) = \sigma_{12}/(\sigma_1 \sigma_2)$ satisfies

$$|h(\sigma_1, \sigma_2, \sigma_{12}) - h(\sigma'_1, \sigma'_2, \sigma'_{12})| \le \frac{3 \max\{|\sigma_1 - \sigma'_1|, |\sigma_2 - \sigma'_2|, |\sigma_{12} - \sigma'_{12}|\}}{b^2}.$$

The result then follows from the previous lemma. □

Let $J_\theta$ denote the $2 \times 2$ rotation matrix of angle $\theta$. Let

$$J_\Sigma = \begin{pmatrix} \cos(\theta(\Sigma)) & -\sin(\theta(\Sigma)) \\ \sin(\theta(\Sigma)) & \cos(\theta(\Sigma)) \end{pmatrix} \tag{31}$$

denote the Jacobi rotation where

$$\theta(\Sigma) = \frac{1}{2} \tan^{-1}\left(\frac{2\Sigma_{12}}{\Sigma_{11} - \Sigma_{22}}\right). \tag{32}$$

LEMMA A.3. *Let $F$ be a bivariate distribution with $2 \times 2$ covariance matrix $\Sigma$. Let $J = J_\Sigma$ and $\hat{J} = J_{\hat{\Sigma}}$. Then,*

$$\mathbb{P}(\|\hat{J}^T \hat{\Sigma} \hat{J} - J^T \Sigma J\|_\infty > \epsilon) \leq c_5 p^2 e^{-nc_6 \epsilon^2}. \tag{33}$$

PROOF. Note that $\theta(\Sigma)$ a bounded, uniformly continuous function of $\Sigma$. Similarly, the entries of $J_\theta$ are also bounded, uniformly continuous functions of $\Sigma$. The result then follows from (29). □

For any pair $(\alpha, \beta)$, let $\theta(\alpha, \beta)$ denote the angle of the principal component rotation and let $J(\alpha, \beta, \theta)$ denote the Jacobi rotation on $(\alpha, \beta)$. Define the selection operator

$$\Delta : \mathcal{S}_p \to \{(j, k) : 1 \leq j < k \leq p\}$$

by $\Delta(\Sigma) = (\alpha, \beta)$ where $\rho_{\alpha,\beta} = \arg\max_{ij} \rho_{ij}$. In case of ties, define $\Delta(\Sigma)$ to be the set of pairs $(\alpha, \beta)$ at which the maximum occurs. Hence, $\Delta$ is multivalued on a subset $\mathcal{S}_p^* \subset \mathcal{S}_p$ of measure 0. The one-step treelet operator $T : \mathcal{S}_p \to \mathcal{S}_p$ is defined by

$$T(\Sigma) = \{J^T \Sigma J : J = J(\alpha, \beta, \theta(\alpha, \beta)), (\alpha, \beta) \in \Delta(\Sigma)\}. \tag{34}$$

Formally, $T$ is a multivalued map because of potential ties.

PROOF OF THEOREM 1. The proof is immediate from the lemmas. For the matrices $\hat{\Sigma}_n$, we have that $\|\hat{\Sigma}_n - \Sigma\|_\infty < \delta_n$ except on a set $A_n^c$ of probability tending to 0 at rate $O(n^{-(K-2c)})$. Hence, on the set $A_n = \{\hat{\Sigma}_n : \|\hat{\Sigma}_{n,b}^* - \hat{\Sigma}_n\|_\infty < \delta_n\}$, we have that $T(\hat{\Sigma}_n) \in \mathcal{T}_n(\Sigma)$. The same holds at each step. □

**A.2. Proof of Lemma 1.** Consider first the case where at each level in the tree the treelet operator combines a coarse-grained variable with a singleton according to $\{\{x_1, x_2\}, x_3\}, \ldots$. Let $s_0 = x_1$. For $\ell = 1$, the $2 \times 2$ covariance submatrix $\Sigma^{(0)} \equiv \mathbb{V}\{(s_0, x_2)\} = \sigma_1^2 \begin{pmatrix} 1 & 1 \\ 1 & 1 \end{pmatrix}$. A principal component analysis of $\Sigma^{(0)}$ gives $\theta_1 = \pi/4$ and $s_1 = \frac{1}{\sqrt{2}}(x_1 + x_2)$. By induction, for $1 \leq \ell \leq p-1$, $\Sigma^{(\ell-1)} \equiv \mathbb{V}\{(s_{\ell-1}, x_{\ell+1})\} = \sigma_1^2 \begin{pmatrix} \ell & \sqrt{\ell} \\ \sqrt{\ell} & 1 \end{pmatrix}$. PCA on $\Sigma^{(\ell-1)}$ gives the (unconstrained) rotation angle $\theta_\ell = \arctan\sqrt{\ell}$, and the new sum variable $s_\ell = \frac{1}{\sqrt{\ell+1}} \sum_{i=1}^{\ell+1} x_i$.

More generally, at level $\ell$ of the tree, the treelet operator combines two sum variables $u = \frac{1}{\sqrt{m}} \sum_{i \in \mathcal{A}_u} x_i$ and $v = \frac{1}{\sqrt{n}} \sum_{j \in \mathcal{A}_v} x_j$, where $\mathcal{A}_u, \mathcal{A}_v \subseteq \{1, \ldots, p\}$ denote two disjoint index subsets with $m = |\mathcal{A}_u|$ and $n = |\mathcal{A}_v|$ number of terms, respectively. The $2 \times 2$ covariance submatrix

$$(35) \quad \Sigma^{(\ell-1)} \equiv \mathbb{V}\{(u,v)\} = \sigma_1^2 \begin{pmatrix} m & \sqrt{mn} \\ \sqrt{mn} & n \end{pmatrix}.$$

The correlation coefficient $\rho_{uv} = 1$ for any pair $(u,v)$; thus, the treelet operator $T_\ell$ is a multivariate function of $\Sigma$. A principal component analysis of $\Sigma^{(\ell-1)}$ gives the eigenvalues $\lambda_1 = m + n$, $\lambda_2 = 0$, and eigenvectors $e_1 = \frac{1}{\sqrt{m+n}}(\sqrt{m}, \sqrt{n})^T$, $e_2 = \frac{1}{\sqrt{m+n}}(-\sqrt{n}, \sqrt{m})^T$. The rotation angle

$$(36) \quad \theta_\ell = \arctan\sqrt{\frac{n}{m}}.$$

The new sum and difference variables at level $\ell$ are given by

$$(37) \quad \begin{aligned} s_\ell &= \frac{1}{\sqrt{m+n}}(+\sqrt{m}u + \sqrt{n}v) \\ &= \frac{1}{\sqrt{m+n}} \sum_{i \in \{\mathcal{A}_u, \mathcal{A}_v\}} x_i, \\ d_\ell &= \frac{1}{\sqrt{m+n}}(-\sqrt{n}u + \sqrt{m}v) \\ &= \frac{1}{\sqrt{m+n}}\left(-\sqrt{\frac{n}{m}}\sum_{i \in \mathcal{A}_u} x_i + \sqrt{\frac{m}{n}}\sum_{j \in \mathcal{A}_v} x_j\right). \end{aligned}$$

The results of the lemma follow.

**A.3. Proof of Theorem 2.** Assume that variables from different blocks have not been merged for levels $\ell' < \ell$, where $1 \leq \ell \leq p$. From Lemma 1, we then know that any two sum variables at the preceding level $\ell - 1$ have the general form $u = \frac{1}{\sqrt{m}} \sum_{i \in \mathcal{A}_u} x_i$ and $v = \frac{1}{\sqrt{n}} \sum_{j \in \mathcal{A}_v} x_j$, where $\mathcal{A}_u$ and $\mathcal{A}_v$ are two disjoint index subsets with $m = |\mathcal{A}_u|$ and $n = |\mathcal{A}_v|$ number of terms, respectively. Let $\delta_k = \sigma/\sigma_k$.

If $\mathcal{A}_u \subseteq \mathcal{B}_i$ and $\mathcal{A}_v \subseteq \mathcal{B}_j$, where $i \neq j$, that is, the subsets belong to *different* blocks, then

$$(38) \quad \Sigma^{(\ell-1)} = \mathbb{V}\{(u,v)\} = \begin{pmatrix} m\sigma_i^2 & \sqrt{mn}\sigma_{ij} \\ \sqrt{mn}\sigma_{ij} & n\sigma_j^2 \end{pmatrix} + \sigma^2 I.$$

The corresponding "between-block" correlation coefficient

$$(39) \quad \rho_B^{(\ell-1)} = \frac{\sigma_{ij}}{\sigma_i \sigma_j} \frac{\sqrt{mn}}{\sqrt{m + \delta_i^2}\sqrt{n + \delta_j^2}} \leq \frac{\sigma_{ij}}{\sigma_i \sigma_j}$$

with equality ("worst-case scenario") if and only if $\sigma = 0$.

If $\mathcal{A}_u, \mathcal{A}_v \subset \mathcal{B}_k$, that is, the subsets belong to the *same block*, then

$$\text{(40)} \quad \Sigma^{(\ell-1)} = \mathbb{V}\{(u,v)\} = \sigma_k^2 \begin{pmatrix} m & \sqrt{mn} \\ \sqrt{mn} & n \end{pmatrix} + \sigma^2 I.$$

The corresponding "within-block" correlation coefficient

$$\text{(41)} \quad \rho_W^{(\ell-1)} = \frac{1}{\sqrt{1 + (m+n)/(mn)\delta_k^2 + (1/(mn))\delta_k^4}}$$
$$\geq \frac{1}{\sqrt{1 + 3\max(\delta_k^2, \delta_k^4)}},$$

with the "worst-case scenario" occurring when $m = n = 1$, that is, when singletons are combined. Finally, the main result of the theorem follows from the bounds in Equations (39) and (41), and the fact that

$$\text{(42)} \quad \max \rho_B^{(\ell-1)} < \min \rho_W^{(\ell-1)}$$

for $\ell = 1, 2, \ldots, p - K$ is a sufficient condition for not combining variables from different blocks. If the inequality equation (13) is satisfied, then the coefficients in the treelet expansion have the general form in equation (37) at any level $\ell$ of the tree. With white noise added, the expansion coefficients have variances $\mathbb{V}\{s_\ell\} = (m+n)\sigma_k^2 + \sigma^2$ and $\mathbb{V}\{d_\ell\} = \sigma^2 \frac{m^2+n^2}{mn(m+n)}$. Furthermore, $\mathbb{E}\{s_\ell\} = \mathbb{E}\{d_\ell\} = 0$.

**Acknowledgments.** We are grateful to R. R. Coifman and P. Bickel for helpful discussions. We would also like to thank K. Roeder and three anonymous referees for comments that improved the manuscript, Dr. D. Rimm for sharing his database on hyperspectral images, and Dr. G. L. Davis for contributing his expertise on the histology of the tissue samples.

TREELETS 39Zou, H. and Hastie, T. (2005). Regularization and variable selection via the elastic net. *J. Roy. Statist. Soc. Ser. B* **67** 301–320. MR2137327

Zou, H., Hastie, T. and Tibshirani, R. (2006). Sparse principal component analysis. *J. Comput. Graph. Statist.* **15** 265–286. MR2252527
A. B. Lee
L. Wasserman
Department of Statistics
Carnegie Mellon University
Pittsburgh, Pennsylvania
USA
E-mail: annlee@stat.cmu.edu
    larry@stat.cmu.edu

B. Nadler
Department of Computer Science
    and Applied Mathematics
Weizmann Institute of Science
Rehovot
Israel
E-mail: boaz.nadler@weizmann.ac.il